\documentclass[10pt,twocolumn]{article}
\usepackage[margin=0.60in,columnsep=0.22in]{geometry}
\usepackage{amsmath,amssymb,amsthm,mathtools,microtype,enumitem}
\usepackage{booktabs,tabularx}
\usepackage{graphicx}
\usepackage{tikz,standalone}
\usetikzlibrary{arrows.meta,patterns,calc}
\usepackage[hidelinks]{hyperref}
\usepackage{siunitx}

\newtheorem{theorem}{Theorem}
\newtheorem{proposition}{Proposition}
\newtheorem{corollary}{Corollary}

\newtheorem{definition}{Definition}
\newtheorem{assumption}{Assumption}
\theoremstyle{remark}
\newtheorem*{remark}{Remark}

\newcommand{\spow}[2]{\left\lfloor #1\right\rceil^{#2}}
\newcommand{\R}{\mathbb R}
\DeclareMathOperator{\sgn}{sgn}

\newcommand{\velprofileso}{f}
\newcommand{\velenergyso}{F}
\newcommand{\feedbackmapso}{v}

\newcommand{\velerrorso}{\sigma}
\newcommand{\restoringso}{\phi}
\newcommand{\dampingso}{\psi}
\newcommand{\potentialso}{U}

\newcommand{\velprofilefo}{g}
\newcommand{\velenergyfo}{G}
\newcommand{\feedbackmapfo}{\nu}
\newcommand{\combinedmapfo}{K}
\newcommand{\potentialfo}{P}

\newcommand{\nominalacc}{k}
\newcommand{\velestimate}{\eta}
\newcommand{\velestimateerror}{y}
\newcommand{\distestimate}{\zeta}
\newcommand{\injectionbound}{\Gamma}
\newcommand{\injectionshape}{\omega}

\hypersetup{
  pdfauthor={Christian Fredrik Sætre},
  pdftitle={Profile-Anchored Control: Velocity Shaping with Exact Disturbance Rejection and Transient-Free Replanning},
}

\title{Profile-Anchored Control: Velocity Shaping with Exact Disturbance Rejection and Transient-Free Replanning}

\author{Christian Fredrik Sætre\footnotemark[1] \and Morten Dinhoff Pedersen%
\thanks{The authors are with the Department of Engineering
Cybernetics, Norwegian University of Science and Technology,
NO-7491 Trondheim, Norway (e-mail: \texttt{christian.f.satre@ntnu.no};
\texttt{morten.d.pedersen@ntnu.no}).}}
\date{November 19, 2026}

\begin{document}
\maketitle
\vspace{-2em}

\begin{abstract}
In many motion-control problems, how a target is approached matters as much as whether it is reached: speed and deceleration must stay within actuator, safety or comfort limits despite unknown loads and disturbances. Position-dependent velocity profiles provide a natural way to describe such requirements. We show that the square of such a profile can be used both to plan the motion and to construct the controller that executes it. Our main result, profile-anchored control with exact rejection (PACER), recovers the nominal motion after a finite time under an unknown gain and load, and then allows the profile to be replaced mid-motion without a transient. It combines a static law for scalar second-order systems with a dynamic law for first-order ones, used for disturbance rejection. The static law guarantees uniform global asymptotic stability under an unknown input gain with a known lower bound, and finite-time stability under local homogeneity, while the first-order law gives analogous  guarantees under disturbances of bounded rate and provides the exact rejection in PACER. Replanning of a profile under speed, acceleration and jerk limits can be formulated as a convex program. The approach is illustrated on a simulated elevator.

\end{abstract}


\section{Introduction}

In stabilization problems, the specification is usually the destination:
an equilibrium to be reached, e.g., asymptotically or in finite time, while
the route towards it is shaped only indirectly through the choice of
feedback gains.

In motion control, however, the route is often part of the
specification itself. Think, for example, of a train approaching a
station, an elevator approaching a floor or a robot placing an object:
each must respect limits on speed and deceleration, set by actuators,
safety or comfort, and reach the target precisely despite unknown
loads. 

Such requirements are naturally expressed by prescribing the
velocity as a function of the remaining distance. This is common in
servo design, although normally  with a fixed profile and loop
(see, e.g., \cite{workman1987adaptive,venkataramanan2002new}).
Moreover, since such a profile depends on position rather than time,
the motion can resume it from the current state after a disturbance,
instead of catching up with a time-parameterized reference
\cite{hauser1995maneuver}. Building such requirements into the
feedback thus shapes the transient itself, not only the convergence,
without a separate trajectory planner or safety filter.

The idea is simplest for a double integrator $\ddot{q}=u$, with $q$
the position relative to the target and $u$ the control input. A specification is then a continuous velocity profile
$\dot{q}=\velprofileso(q)$ satisfying $q\velprofileso(q)<0$ away from
$q=0$. Here, speed limits correspond to
bounded profiles, while non-monotone profiles permit alterations to the
velocity along the way. The deviation from the profile corresponds to
the transverse variable $\velerrorso:=\velprofileso(q)-\dot{q}$, which,
where $\velprofileso'$ exists, satisfies
\begin{equation*}
 \dot{\velerrorso}=\velprofileso'(q)\dot{q}-u.
\end{equation*}

In sliding-mode control \cite{shtessel2014sliding} and backstepping
control  \cite{krstic1995nonlinear}, one commonly treats such a profile as
either a sliding manifold or a virtual control, respectively, and
imposes a desired transverse dynamics by cancelling the term
$\velprofileso'(q)\dot{q}$. This ties the feedback to $\velprofileso'$,
which is singular for certain terminal profiles \cite{wu1998terminal}, and has
motivated reparameterizations of the sliding variable, such as
nonsingular terminal sliding modes~\cite{feng2002non}.

In the spirit of Utkin's equivalent-control viewpoint~\cite{utkin1992sliding}, we here instead simply replace the cancellation of
$\velprofileso'(q)\dot{q}$ with the nominal control required on the
profile $\dot{q}=\velprofileso(q)$:
\begin{equation*}
 u_{\rm nom}=\velenergyso'(q), \qquad \velenergyso(q):=\tfrac12\velprofileso(q)^2,
\end{equation*}
so that
$\velenergyso'(q)=\velprofileso'(q)\velprofileso(q)$ wherever
$\velprofileso'$ exists. Hence $\velenergyso$ can be continuously
differentiable even when $\velprofileso'$ is discontinuous or does not
exist at the target, so that terminal profiles are treated directly, and the global
profile can be reshaped without altering its local behaviour.

Taking inspiration from orbital stabilization, where a nominal motion
along a target curve is combined with feedback on coordinates
transverse to it \cite{shiriaev2008can,saetre2022orbital,saetre2023orbital},
we then
complement the nominal control with a feedback map $\feedbackmapso$:
\begin{equation*}
 u=\velenergyso'(q)+\feedbackmapso(q,\velerrorso).
\end{equation*}
This allows for flexibility in both the profile and feedback map design, with  the profile $\velprofileso$ describing the nominal motion, and the
feedback map $\feedbackmapso$ specifying how deviations from it are corrected.
Thus the profile can be chosen from the
task requirements, such as limits on speed and deceleration, while the
feedback map can be chosen to handle, for example, unknown loads and disturbances. 

However, a static feedback map cannot reject perturbations that
persist on the profile, such as gravity or friction offsets, and
under an unknown gain the motion follows the profile only
approximately. Our main contribution is to combine a static law of this form
with a dynamic controller from its first-order counterpart. The result is 
PACER, short for profile-anchored control with exact rejection,
under which the nominal motion can be recovered exactly after a finite
time. Since the profile is parameterized by position, it can moreover be
replaced along the way by a new profile through the current state,
with no transient and no reset of the disturbance estimate; and,
taking inspiration from \cite{verscheure2009time,lipp2014minimum},
replanning under speed, acceleration and jerk limits is convex in
$\velenergyso$, with the planner's variable and the controller's
nominal input being the same object.

 To the best of our
knowledge, PACER is the first design in which the nominal motion
toward a position-dependent velocity profile is recovered exactly,
in finite time, under an unknown gain and time-varying loads, and in
which the profile can be replaced along the way with a proven
transient-free switch. In detail, our contributions (listed by
section number) are:
\begin{enumerate}
 \item[\ref{sec:so}:] A static law for second-order systems that
 is uniformly globally asymptotically stable (UGAS) under an unknown
 gain and perturbations, and finite-time
 stable under local homogeneity (Thm.~\ref{thm:general-f}); under
 further conditions the profile bounds the velocity
 (Cor.~\ref{cor:envelope}).
 \item[\ref{sec:fo}:] The same construction for first-order
 systems, giving a dynamic law with integral action that contains PI
 and super-twisting control and rejects disturbances of bounded rate
 (Prop.~\ref{prop:first-order}).
 \item[\ref{sec:so-dyn}:] Their combination, PACER, which, for
 locally Lipschitz nominal laws, recovers (semiglobally) the nominal dynamics
 exactly after a finite time under an unknown gain
 and disturbances of bounded rate (Thm.~\ref{thm:so-fo}), and keeps
 the velocity envelope up to a margin that can be made arbitrarily
  small (Cor.~\ref{cor:pace-envelope}).  Both conclusions also hold for
 profiles with finite-time convergence
 for initial states away from the target
 (Prop.~\ref{prop:so-fo-terminal}).
 \item[\ref{sec:replanning}:] Replanning
 and switching to a new profile without transient, where the new profile is followed exactly from the
 switch (Prop.~\ref{prop:replanning}), and planning it under speed,
 acceleration and jerk limits is a convex problem in $\velenergyso$.
\end{enumerate}
The results are illustrated on an elevator in Section~\ref{sec:example},
and the paper closes with a discussion, including related work.

\section{Static feedback for second-order systems}\label{sec:so}
Consider a second-order system
\begin{equation}
 \ddot{q}(t)=bu(t)+d(t,q(t),\dot{q}(t)), \quad q(t)\in\R,
 \label{eq:second-order}
\end{equation}
where $d$ is a continuous, possibly time-varying perturbation, and $b\in[\underline{b},\infty)$ is an unknown constant with known lower bound
$\underline{b}>0$. We omit time arguments in the following  unless needed for clarity.

\begin{remark}For the sake of simplicity, we retain the nominal reference $b=1$
from the Introduction in the following control synthesis, even though it lies outside the admissible
gain range when $\underline{b}>1$; this does not affect the results, and there is no loss of generality.
If a good estimate $b_0>0$ is available, however, scaling the implemented
control by $1/b_0$ gives the normalized gain $b/b_0$, with nominal
value one.
\end{remark}

The objective is to drive the velocity $\dot{q}$ onto a velocity profile $q\mapsto\velprofileso(q)$ from the following class:
\begin{definition}
\label{def:velocity-profile}
A profile $\velprofileso:\R\to\R$ is admissible if:
\begin{enumerate}
     \item  $\velprofileso$  is $\mathcal C^0$ (continuity); 
\item $\velprofileso(0)=0$  (zero-crossing);
 \item $q\velprofileso(q)<0$ for $q\neq0$ (motion towards the target); 
 \item $\velenergyso(q):=\tfrac12\velprofileso(q)^2$ is $\mathcal C^1$ (continuous nominal control).
\end{enumerate}
\end{definition}
The problem of bringing $\dot{q}$ onto the curve traced out by $\velprofileso$ corresponds to driving the following transverse variable (not necessarily monotonically) to zero:
\begin{equation*}
     \velerrorso:=\velprofileso(q)-\dot{q}.
\end{equation*}
Therefore, under the assumption that $b=1$ and $d=0$, the nominal control on $\dot{\velerrorso}=\velerrorso=0$ must be given by $u_{\text{nom}}=\velenergyso'(q)$ (that is,  $=\velprofileso'(q)\velprofileso(q)$ whenever $\velprofileso'$ exists), which is continuous by the fourth property of $\velprofileso$ above.

We suggest a control law that augments the nominal controller by a suitable scalar function $(q,\velerrorso)\mapsto\feedbackmapso(q,\velerrorso)$:
\begin{equation}
 u=\velenergyso'(q)+\feedbackmapso(q,\velerrorso).
 \label{eq:general-controller}
\end{equation}
The feedback map is taken from the following class:
\begin{definition}
    \label{def:feedbackmap}
    For an admissible profile $\velprofileso$ according to Def.~\ref{def:velocity-profile}, a corresponding feedback map $\feedbackmapso:\R^2\to\R$ satisfies:
\begin{enumerate}
 \renewcommand{\labelenumi}{\roman{enumi}.}
 \item $\feedbackmapso$ is $\mathcal{C}^0$;
 \item  $\feedbackmapso(q,0)=0$;
 \item $(s-\velprofileso(q))\bigl[\feedbackmapso(q,s)-\feedbackmapso(q,\velprofileso(q))\bigr]>0$
 for $s\neq \velprofileso(q)$; 
 \item $q\bigl[\feedbackmapso(q,\velprofileso(q))+\velenergyso'(q)\bigr]<0$
 for $q\neq0$;
 \item $\int_0^q\feedbackmapso(r,\velprofileso(r))\,dr+\velenergyso(q)\to-\infty$
 as $|q|\to\infty$.
\end{enumerate}
\end{definition}

These conditions have a direct mechanical interpretation: 
The controller \eqref{eq:general-controller} can be equivalently written as
\begin{equation}
    u=-\restoringso(q)-\dampingso(q,\dot{q}),
    \label{eq:springdamper}
\end{equation}
where $\restoringso$  and  $\dampingso$ can be considered as restoring and dissipative terms, respectively, and are defined as
\begin{subequations}
    \label{eq:phipsi-def}
    \begin{align}
    \restoringso(q)&:=-\velenergyso'(q)-\feedbackmapso(q,\velprofileso(q)),
    \\
    \dampingso(q,z)&:=\feedbackmapso(q,\velprofileso(q))-\feedbackmapso(q,\velprofileso(q)-z).
    \label{eq:phipsi}
    \end{align}
\end{subequations}
Indeed, Def.~\ref{def:feedbackmap} ensures $q\restoringso(q)>0$ for $q\neq0$ and
$z\dampingso(q,z)>0$ for $z\neq0$. Thus $\restoringso$ is restoring and $\dampingso$ is
dissipative, while condition v makes the associated restoring potential
radially unbounded.  Condition~\emph{ii} also makes $\dot{q}=\velprofileso(q)$ a solution of the nominal loop $b=1$, $d=0$.
Although the spring--damper structure \eqref{eq:springdamper} of course is well
known from energy- and passivity-based control
methods and nonlinear damping designs
\cite{ortega2002interconnection,van2017l2,ruderman2021optimal}, such designs are commonly not built around a nominal motion. Here, however, the velocity profile $\velprofileso$ (Def.~\ref{def:velocity-profile}) specifies the
approach, and the feedback map $\feedbackmapso$ (Def.~\ref{def:feedbackmap}) the restoring and
dissipative action around it.\footnote{The identity $\velenergyso'=\velprofileso'\velprofileso$
is standard in path-velocity planning, where
$\velenergyso$ is the decision variable
\cite{verscheure2009time,lipp2014minimum}; see also
Section~\ref{sec:replanning}. What is new is its use as the only
model-dependent term of the feedback, together with a general
$\feedbackmapso$.}

While the control law is derived from the desire to bring the velocity onto the curve traced out by $\velprofileso$, it cannot in general guarantee that $\velerrorso=0$ is made forward invariant due to the uncertainty in $b$ and the perturbation. The following nevertheless provides conditions ensuring global asymptotic or finite-time stability of the origin:
\begin{theorem}
\label{thm:general-f}
Consider the system \eqref{eq:second-order} under the control law \eqref{eq:general-controller}, with $\velprofileso$ and $\feedbackmapso$ satisfying Defs.~\ref{def:velocity-profile} and \ref{def:feedbackmap}.

For some $\delta\in[0,\underline{b})$, suppose the  perturbation satisfies
\begin{equation}
 |d(t,q,\dot{q})|\le\delta\bigl|\feedbackmapso(q,\velprofileso(q)-\dot{q})-\feedbackmapso(q,\velprofileso(q))\bigr|
 \label{eq:incremental-bound}
\end{equation}
for all $t\ge 0$.
 Then the origin, $(q,\dot{q})=(0,0)$, of \eqref{eq:second-order}
is uniformly globally asymptotically stable under \eqref{eq:general-controller}.

If, in addition, there exists an $\alpha\in(1/2,1)$ and a neighborhood
$\mathcal N\subset\R^2$ of the origin such that, for all
$(q,s)\in\mathcal N$ and $\ell>0$ for which
$(\ell q,\ell^\alpha s)\in\mathcal N$, the homogeneity conditions
\begin{equation}
 \velprofileso(\ell q)=\ell^\alpha \velprofileso(q)\quad\text{and}\quad
 \feedbackmapso(\ell q,\ell^\alpha s)
 =\ell^{2\alpha-1}\feedbackmapso(q,s)
 \label{eq:general-homogeneity}
\end{equation}
hold, then the origin is globally finite-time stable.
\end{theorem}

\begin{proof}
In terms of \eqref{eq:phipsi-def}, the closed loop is
$\ddot{q}=-b\restoringso(q)-b\dampingso(q,\dot{q})+d$ with
$q\restoringso(q)>0$ and $\dot{q}\dampingso(q,\dot{q})>0$ off zero, and
\eqref{eq:incremental-bound} reads $|d|\le\delta|\dampingso(q,\dot{q})|$.
We i) show UGAS by a Matrosov argument and ii) strictify the Lyapunov
function near the origin to obtain finite-time stability.

i) By ii, iv and v of Def.~\ref{def:feedbackmap}, the potential
$\potentialso(q):=\int_0^q\restoringso(r)\,dr$ is positive definite and
radially unbounded, and so is
$V_0(q,\dot{q}):=\tfrac12\dot{q}^2+b\potentialso(q)$, with
\begin{equation*}
 \dot{V}_0=-b\dot{q}\dampingso(q,\dot{q})+\dot{q}d
 \le-c_0|\dot{q}|\,|\dampingso(q,\dot{q})|,
\end{equation*}
where $c_0:=\underline{b}-\delta>0$.
On $\dot{q}=0$, $d=0$ by \eqref{eq:incremental-bound}, so
$M(q,\dot{q}):=q\dot{q}$ has $\dot{M}=-bq\restoringso(q)<0$ for
$q\neq0$, and a neste Matrosov theorem \cite{loria2005nested} gives UGAS.

ii) Now, let $z(q):=\sgn(q)\sqrt{2\potentialso(q)}$, so that
$z(q)z'(q)=\restoringso(q)$ for $q\neq0$, and take the new candidate
\begin{equation*}
 V_\varepsilon:=V_0+\varepsilon\Psi
\end{equation*}
with 
\begin{equation*}
     \Psi(q,\dot{q}):=
 \begin{cases}
 \dfrac{z(q)^3\dot{q}}{z(q)^2+\dot{q}^2}, & (q,\dot{q})\neq(0,0),\\[6pt]
 0, & (q,\dot{q})=(0,0),
 \end{cases}
\end{equation*}
and  where $\varepsilon\in(0,b)$ is to be chosen \cite{mazenc2004strong,cruz2018strict}.
By Young's inequality,
$|\Psi|\le\tfrac12z^2$, $|\partial\Psi/\partial\dot{q}|\le|z|$ and
$|\partial\Psi/\partial q|\le\tfrac32|\restoringso(q)|$, so
$V_\varepsilon$ is $\mathcal{C}^1$, with
$V_\varepsilon\ge\tfrac12\dot{q}^2+\tfrac{b-\varepsilon}{2}z^2$ positive
definite and  radially unbounded. Moreover,
\begin{equation*}
 \dot{V}_\varepsilon\le-c_0|\dot{q}|\,|\dampingso(q,\dot{q})|+\varepsilon R(q,\dot{q}),
\end{equation*}
where
\begin{equation*}
 R(q,\dot{q}):=\frac{\partial\Psi}{\partial q}\,\dot{q}
 -b\,\frac{\partial\Psi}{\partial\dot{q}}\bigl[\restoringso(q)+\dampingso(q,\dot{q})\bigr]
 +\delta\Bigl|\frac{\partial\Psi}{\partial\dot{q}}\Bigr|\,|\dampingso(q,\dot{q})|
\end{equation*}
satisfies $R(q,0)=-bz(q)\restoringso(q)<0$ for $q\neq0$.

To ensure $\dot{V}_\varepsilon<0$ also away from $\dot{q}=0$, we use
\eqref{eq:general-homogeneity}. Let $\mathcal N$ be a neighborhood of
the origin on which it holds, taken as an open box, hence invariant
under $(q,s)\mapsto(\ell q,\ell^\alpha s)$ for $0<\ell\le1$
($\mathcal N=\R^2$ if it holds globally). By continuity, choose
$\bar{c}>0$ such that, on
$\Omega_{\bar{c}}:=\{(q,\dot{q}):V_0(q,\dot{q})\le\bar{c}\}$, both
$(q,\velprofileso(q))$ and $(q,\velprofileso(q)-\dot{q})$ lie in
$\mathcal N$; the box shape keeps them there under inward dilation.
On $\Omega_{\bar{c}}$, $\restoringso$, $\potentialso$ and $\dampingso$
are thus homogeneous of degrees $2\alpha-1$, $2\alpha$ and $2\alpha-1$
under $(q,\dot{q})\mapsto(\ell q,\ell^\alpha\dot{q})$, $0<\ell\le1$, so
that $z(\ell q)=\ell^\alpha z(q)$, and $V_0$, $\Psi$ and hence
$V_\varepsilon$ are homogeneous of degree $2\alpha$, while every term of
\begin{equation*}
 B(q,\dot{q}):=-c_0|\dot{q}|\,|\dampingso(q,\dot{q})|+\varepsilon R(q,\dot{q})
\end{equation*}
has degree $3\alpha-1$.
On the compact level set $\{V_0=\bar{c}\}$, choose by continuity an open
neighborhood $\mathcal U$ of its intersection with $\{\dot{q}=0\}$ where
$R<0$, and let $\mathcal K:=\{V_0=\bar{c}\}\setminus\mathcal U$. Then
\begin{equation*}
 \begin{aligned}
 m&:=\min_{\mathcal K}|\dot{q}|\,|\dampingso(q,\dot{q})|>0,\\
 M&:=\max_{\mathcal K}|R|<\infty,
 \end{aligned}
\end{equation*}
and any $0<\varepsilon<\min\{b,c_0m/(2M)\}$ (the second bound omitted
if $M=0$) gives $\dot{V}_\varepsilon\le B<0$ on $\{V_0=\bar{c}\}$. With
\begin{equation*}
 \theta:=\frac{3\alpha-1}{2\alpha}\in\Bigl(\tfrac12,1\Bigr),
\end{equation*}
$B$ and $V_\varepsilon^\theta$ have the same degree, so
\begin{equation*}
 c:=\min_{\{V_0=\bar{c}\}}\bigl(-B/V_\varepsilon^\theta\bigr)>0,
\end{equation*}
and inward dilation gives, on $\Omega_{\bar{c}}\setminus\{0\}$,
\begin{equation*}
 \dot{V}_\varepsilon\le B\le-cV_\varepsilon^\theta
\end{equation*}
\cite{bhat2005geometric}. Since $\dot{V}_0\le0$, $\Omega_{\bar{c}}$ is
forward invariant, and UGAS gives a finite entry time $t_{\rm e}$; as
$\theta<1$, the origin is then reached no later than
$t_{\rm e}+V_\varepsilon(q(t_{\rm e}),\dot{q}(t_{\rm e}))^{1-\theta}/(c(1-\theta))$.
Hence the origin is globally finite-time stable.
\end{proof}

Theorem~\ref{thm:general-f} contains several familiar controllers as
special cases. The perhaps most familiar is obtained by selecting
\begin{equation*}
    \velprofileso(q)=-\kappa q
    \quad\text{and}\quad
    \feedbackmapso(q,s)=\beta s,
\end{equation*}
with $\beta>\kappa>0$, which gives a linear PD controller, over or
critically damped for the nominal gain $b=1$. Another choice, namely
\begin{equation*}
    \velprofileso(q)=-\kappa\spow{q}{\alpha},
    \qquad
    \feedbackmapso(q,s)=\beta\spow{s}{\frac{2\alpha-1}{\alpha}},
\end{equation*}
with $\alpha\in(\frac{1}{2},1)$, where $\spow{x}{a}:=|x|^a\sgn(x)$, gives
\begin{equation*}
    u
    =
    \alpha\kappa^2\spow{q}{2\alpha-1}
    +\beta\spow{\velerrorso}{\frac{2\alpha-1}{\alpha}},
    \qquad
    \velerrorso=-\kappa\spow{q}{\alpha}-\dot{q},
\end{equation*}
for which the restoring condition reduces to
$\beta>\alpha\kappa^{1/\alpha}$ and \eqref{eq:general-homogeneity}
holds. This family of terminal sliding-mode controllers was previously
proposed in \cite{saetre2022orbital} and experimentally tested in
\cite{sandvand2025control}, and has natural extensions (not considered
here) including adding integral action à la \cite{kamal2016continuous},
sector bounds à la \cite{ruderman2021optimal}, and a twisting-like
controller obtained at the boundary $\alpha=\frac{1}{2}$. With the same
profile but the separable map
$\feedbackmapso(q,s)=\beta\spow{\velprofileso(q)}{\frac{2\alpha-1}{\alpha}}-\beta\spow{\velprofileso(q)-s}{\frac{2\alpha-1}{\alpha}}$,
one instead obtains
$u=-\beta\spow{\dot{q}}{\frac{2\alpha-1}{\alpha}}-(\beta\kappa^{\frac{2\alpha-1}{\alpha}}-\alpha\kappa^2)\spow{q}{2\alpha-1}$,
which for the nominal gain is the continuous finite-time controller of
\cite{bhat1998continuous}, under the same restoring condition.

Of main value, however, is that Theorem~\ref{thm:general-f} leaves substantial freedom in the
selection of both the velocity profile $\velprofileso$ and the feedback map
$\feedbackmapso$. In particular, the nominal approach motion can be shaped
through $\velprofileso$, which can, for example, be bounded, saturating or non-monotone. The feedback map $\feedbackmapso$, on the other hand, can be
modified to introduce effects as nonlinear or saturating damping
without changing the selected profile.

Despite its generality, Theorem~\ref{thm:general-f} concerns convergence only. Indeed, since $\velerrorso=0$
need not be invariant for the uncertain system, it does not by itself ensure that a bounded
profile bounds the actual velocity. This, however, can easily be remedied by requiring stronger (sufficient but not necessary) assumptions, which together render the set
\begin{equation}
 \mathcal{S}:=\{(q,\dot{q}):q\dot{q}\le0,\ |\dot{q}|\le|\velprofileso(q)|\}
 \label{eq:envelope}
\end{equation}
 forward invariant. Here, $\mathcal{S}$ can be considered a ``trapping envelope'', with  motions starting at rest ($\dot{q}(0)=0$), or below the profile, neither overshoot the target $q=0$
nor exceed the profile. The following provides such additional conditions:
\begin{corollary}
\label{cor:envelope}
Let the hypotheses of Theorem~\ref{thm:general-f} hold, with
$s\feedbackmapso(q,s)\ge0$ whenever $qs>0$ and $\dot{q}\,d\le0$ whenever
$q\velerrorso>0$. If $b=1$, then $\mathcal{S}$ in \eqref{eq:envelope}
is forward invariant for \eqref{eq:second-order} under
\eqref{eq:general-controller}; while if $b\ge1$ and $\velprofileso$ is
non-increasing, then $\mathcal{S}$ is forward invariant and
$|\dot{q}(t)|\le|\velprofileso(q(0))|$ for all $t\ge0$.
\end{corollary}

\begin{proof}
Let $W:=\tfrac12\dot{q}^2-\velenergyso(q)$, so that
$\mathcal{S}=\{(q,\dot{q}):q\dot{q}\le0,\ W\le0\}$. One has
 $\dot{W}=\dot{q}\bigl[(b-1)\velenergyso'(q)+b\feedbackmapso(q,\velerrorso)+d\bigr]$.
 By  $q\velerrorso>0$, we have for $q>0$ and $\velerrorso>0$, that $\dot{q}<\velprofileso(q)<0$,
$\feedbackmapso(q,\velerrorso)\ge0$ and $d\ge0$, while
$(b-1)\velenergyso'(q)$ vanishes for $b=1$ and is nonnegative for $b\ge1$ with $\velprofileso$ non-increasing,
since then $|\velprofileso|$ is non-decreasing for $q>0$; so $\dot{W}\le0$. The case $q<0$,
$\velerrorso<0$ is symmetric. On the rest axis $\dot{q}=0$, $q\neq0$, one
has $\ddot{q}=-b\restoringso(q)$, since $d(t,q,0)=0$ by
\eqref{eq:incremental-bound}, which has the sign of $-q$, so solutions
cannot leave $\mathcal{S}$ there either, making
 $\mathcal{S}$ forward invariant. 
\end{proof}
The added sign conditions on $\feedbackmapso$ and $d$ in the above
statement ensure the invariance of the envelope
$|\dot{q}|\le|\velprofileso(q)|$, as in non-overshooting designs
\cite{krstic2006nonovershooting,ruderman2025non}, here with no safety
filter (cf.\ control barrier functions \cite{ames2016control}); under
an unknown gain, $b\ge1$ and a non-increasing profile are needed in
addition, since $b>1$ can cause overshoot of the profile wherever it
accelerates.

Note that, for profiles
that accelerate, such as a ramp-up from a creep speed, one requires the
gain mismatch to be rejected, which is what the dynamic extension of
Section~\ref{sec:so-dyn} provides (Corollary~\ref{cor:pace-envelope}).
Before turning to that extension, however, we show that the controller of
Theorem~\ref{thm:general-f}, although tailored to second-order
systems, also yields a dynamic feedback for first-order systems, which
in turn provides the disturbance rejection used in the extension.

\section{Dynamic feedback for first-order systems}\label{sec:fo}
Consider now a first-order system 
\begin{equation}
 \dot{y}(t)=bu(t)+d(t), \quad y(t)\in\R,
 \label{eq:first-order}
\end{equation}
with constant $b\in[\underline{b},\infty)$ and a locally absolutely
continuous disturbance $d:\R_{\ge0}\to\R$. Differentiating \eqref{eq:first-order} yields,
almost everywhere, a second-order system with $\dot{u}$ as the input:
\begin{equation*}
    \ddot{y}=b\dot{u}+\dot{d}.
\end{equation*}
Hence, we can, under certain additional requirements, utilize the scheme of the previous section. To make a clear separation, however, we will here write the desired velocity as $\dot{y}=\velprofilefo(y)$, such that
$\velerrorso=\velprofilefo(y)-\dot{y}$. With this convention and $\velenergyfo:=\tfrac12\velprofilefo^2$,
Theorem~\ref{thm:general-f} suggests a
\emph{dynamic} control law:
\begin{equation*}
  \dot{u}=\velenergyfo'(y)+\feedbackmapso(y,\velprofilefo(y)-\dot{y})  .
\end{equation*}
Yet,  such a controller generally requires knowledge of $\dot{y}$; if
$\dot{y}$ is known, then Theorem~\ref{thm:general-f} can be used
directly, and if both $b$ and $d$ are known sufficiently well, $bu+d$
can serve as a proxy. 

The approach taken here is instead to choose
$\feedbackmapso$ such that $\dot{y}$ enters only through its integral. That is, we impose
\begin{equation*}
    \feedbackmapso(y,s)=\feedbackmapfo'(y)s
\end{equation*}
for some $\feedbackmapfo$ that is $\mathcal{C}^1$ away from $y=0$. Thus,
\begin{equation*}
     \dot{u}=\velenergyfo'(y)+\feedbackmapfo'(y)(\velprofilefo(y)-\dot{y}),
\end{equation*}
which, by integrating $\feedbackmapfo'(y)\dot{y}$, corresponds to 
\begin{subequations}
    \label{eq:fo-dyn-controller}
    \begin{align}
        u&=-\feedbackmapfo(y)+\xi,    
        \\
        \dot{\xi}&=\velenergyfo'(y)+\feedbackmapfo'(y)\velprofilefo(y)=\velprofilefo(y)\left(\velprofilefo'(y)+\feedbackmapfo'(y)\right).
    \end{align}
\end{subequations}
Hence, unlike the control law \eqref{eq:general-controller} which for example requires $\velenergyso'$ to be continuous, the controller \eqref{eq:fo-dyn-controller} only requires continuity of $\feedbackmapfo$ and local integrability in time of $\velprofilefo(y)\left(\velprofilefo'(y)+\feedbackmapfo'(y)\right)$ for it to be continuous. We can therefore relax the conditions of Definitions~\ref{def:velocity-profile} and \ref{def:feedbackmap} somewhat:
\begin{definition}
\label{def:first-order-profiles}
A pair  $\velprofilefo,\feedbackmapfo:\R\to\R$ is admissible if:
\begin{enumerate}
 \renewcommand{\labelenumi}{\alph{enumi}.}
 \item $\velprofilefo$ and $\feedbackmapfo$ are both $\mathcal C^0$ on $\R$ and
 $\mathcal C^1$ on $\R\setminus\{0\}$;
 \item $\velprofilefo(0)=\feedbackmapfo(0)=0$;
 \item $y\velprofilefo(y)<0$ for $y\neq0$;
\item $\liminf_{y\to r}\bigl[\velprofilefo'(y)+\feedbackmapfo'(y)\bigr]>0$ for all $r\in\R$;
  \item 
 $\limsup_{y\to0}\bigl|\velprofilefo(y)\bigl[\velprofilefo'(y)+\feedbackmapfo'(y)\bigr]\bigr|<\infty$;

 \item $\int_0^y \velprofilefo(r)\bigl[\velprofilefo'(r)+\feedbackmapfo'(r)\bigr]\,dr\to-\infty$
 as $|y|\to\infty$.
\end{enumerate}
\end{definition}

Note that, with $\feedbackmapso(y,s)=\feedbackmapfo'(y)s$, condition~\emph{ii} of
Def.~\ref{def:feedbackmap} holds trivially, \emph{iv} and \emph{v}
become \emph{d} and \emph{f}, and \emph{iii} reads
$\feedbackmapfo'>0$ (which Prop.~\ref{prop:first-order} below
does not need); condition~\emph{e} replaces the continuity of
$\velenergyso'$ in Def.~\ref{def:velocity-profile}.
Note also that, at $y=0$, where $\velprofilefo(\velprofilefo'+\feedbackmapfo')$ may be
discontinuous, \eqref{eq:fo-dyn-controller} is understood in the
Filippov sense \cite{filippov1988differential}.

The above conditions are related to several familiar controllers, for example: for some $\mu>\kappa>0$,
choosing $\velprofilefo(y)=-\kappa y$ and $\feedbackmapfo(y)=\mu y$ results in a PI controller:
\begin{equation*}
 u=-\mu y+\xi,\qquad \dot{\xi}=-\kappa(\mu-\kappa)y.
\end{equation*}
The pair
$\velprofilefo(y)=-\kappa\spow{y}{1/2}$ and $\feedbackmapfo(y)=\mu\spow{y}{1/2}$, on the other hand,
instead gives the super-twisting algorithm (STA)
\cite{levant1993sliding,seeber2017stability}:\footnote{Replacing $\spow{y}{1/2}$ in both $\velprofilefo$ and $\feedbackmapfo$ by
$\phi(y)=\spow{y}{1/2}+a\spow{y}{3/2}$, $a>0$, gives an instance of the disturbance-tailored STA
\cite{haimovich2019disturbance}:
$ u=-\mu\phi(y)+\xi$ with 
 $\dot{\xi}=-\kappa(\mu-\kappa)\phi(y)\phi'(y)$.}
\begin{equation*}
 u=-\mu\spow{y}{1/2}+\xi,\qquad
 \dot{\xi}=-\frac{1}{2}\kappa(\mu-\kappa)\sgn(y).
\end{equation*}
More generally, proportional pairs $\velprofilefo=-\kappa\phi$,
$\feedbackmapfo=\mu\phi$ give $\dot{\xi}=-\kappa(\mu-\kappa)\phi\phi'$, which is
the generalized STA of \cite{moreno2011lyapunov}.

The following may be seen as the counterpart of Theorem~\ref{thm:general-f} for the first-order system, providing additional requirements alongside  Def.~\ref{def:first-order-profiles}, including conditions on the (time-varying) disturbance that can be rejected by  the integral action, rendering $y=0$ globally asymptotically (and possibly finite-time) stable:
\begin{proposition}
\label{prop:first-order}
Consider the system \eqref{eq:first-order} under the continuous
dynamic control law \eqref{eq:fo-dyn-controller}, with
$(\velprofilefo,\feedbackmapfo)$ satisfying Def.~\ref{def:first-order-profiles}. Suppose, for some $\delta$ satisfying
\begin{equation}
 0\le\delta<
 \begin{cases}
 2\underline{b}\sqrt{\underline{b}(1-\underline{b})},
     &0<\underline{b}<\tfrac12,\\
 \underline{b},&\underline{b}\ge\tfrac12,
 \end{cases}
 \label{eq:fo-rate-margin}
\end{equation} 
that $|\dot{d}(t)|\le\delta\injectionbound$ for almost all $t\ge0$, where
\begin{equation}
 \injectionbound:=\inf_{y\in\R}\liminf_{r\to y}
 \bigl|\velprofilefo(r)\bigl(\velprofilefo'(r)+\feedbackmapfo'(r)\bigr)\bigr|.
 \label{eq:fo-rate-bound}
\end{equation}
Then  $(y,\xi+d/b)=(0,0)$ is uniformly globally
asymptotically stable (UGAS).

If, in addition, for some $\mu>\kappa>0$ and $\alpha\in[1/2,1)$,
\begin{equation}
 \lim_{y\to0}\frac{\velprofilefo(y)}{\spow{y}{\alpha}}=-\kappa,\qquad
 \lim_{y\to0}\frac{\velprofilefo'(y)+\feedbackmapfo'(y)}{|y|^{\alpha-1}}=\alpha(\mu-\kappa),
 \label{eq:fo-local-terminal}
\end{equation}
then $(y,\xi+d/b)=(0,0)$ is globally finite-time stable.
\end{proposition}

\begin{proof}
We first show that 1) a Lyapunov function in transformed coordinates is
strictly decreasing on compact sets away from the origin, which gives
UGAS, and then 2) use \eqref{eq:fo-local-terminal} to obtain a finite-time
decay near the origin.

1) Write $\combinedmapfo:=\velprofilefo+\feedbackmapfo$ and set
\begin{equation*}
 w:=\xi+d/b-\combinedmapfo(y),\qquad
 \potentialfo(y):=-\int_0^y \velprofilefo(r)\combinedmapfo'(r)\,dr.
\end{equation*}
Def.~\ref{def:first-order-profiles} gives $\combinedmapfo'>0$ away from zero,
and $\potentialfo$ is positive definite, radially unbounded, and locally Lipschitz.
For $y\neq0$, the transformed dynamics are
\begin{equation*}
 \dot{y}=b(w+\velprofilefo),\qquad
 \dot{w}=\combinedmapfo'[(1-b)\velprofilefo-bw]+\dot{d}/b.
\end{equation*}
Consider the Lyapunov function candidate
\begin{equation}
 V(y,w):=\tfrac12w^2+\varrho \potentialfo(y), \quad \varrho:=\max\{1,b^{-1}-1\},
 \label{eq:fo-strict-function}
\end{equation}
which is positive definite and radially unbounded, with
\begin{equation*}
 \dot{V}
 =\combinedmapfo'\bigl[-bw^2-\varrho b \velprofilefo^2+(1-b-\varrho b)w\velprofilefo\bigr]
+w\dot{d}/b
\end{equation*}
 for $y\neq0$.
For the choice of $\varrho$, condition~\eqref{eq:fo-rate-margin} implies
\begin{equation*}
 |1-b-\varrho b|+\delta/b<2b\sqrt{\varrho}.
\end{equation*}
Indeed, for $b\ge\tfrac12$, this reduces to $\delta<b$, since
$\delta<\underline{b}\le b$. For $b<\tfrac12$, it reduces to
$\delta<2b\sqrt{b(1-b)}$, which follows from
\eqref{eq:fo-rate-margin} because
$r\mapsto2r\sqrt{r(1-r)}$ is increasing on $(0,\tfrac12)$.
Thus $|\dot{d}|\le\delta\injectionbound\le\delta|\velprofilefo(y)\combinedmapfo'(y)|$
for $y\neq0$ gives, for some $c_0>0$ independent of time and the
admissible disturbance,
\begin{equation}
 \dot{V}\le-c_0\combinedmapfo'(y)\bigl[w^2+\velprofilefo(y)^2\bigr]<0
 \qquad(y\neq0).
 \label{eq:fo-strict-decay}
\end{equation}

At a crossing of $y=0$ away from the error origin, $\dot{y}=bw\neq0$,
and $\combinedmapfo'$ is locally integrable since $\combinedmapfo$ is
continuous and increasing; thus $V$ is absolutely continuous along
solutions and \eqref{eq:fo-strict-decay} holds almost everywhere.
Local Filippov solutions exist since the controller has locally bounded
values \cite{filippov1988differential}, and the boundedness of
$\velprofilefo\combinedmapfo'$ near zero (condition~\emph{e}) together with the rate bound give
bounded derivatives of $(y,\xi+d/b)$ on sublevel sets of $V$, hence
forward completeness. By condition~\emph{d} of
Def.~\ref{def:first-order-profiles}, $\combinedmapfo'$ is bounded
below by a positive constant on every bounded set away from zero, and
$w^2+\velprofilefo(y)^2$ is bounded below by a positive constant on
$\{a\le V\le A\}$ for $0<a<A$, since $V=\tfrac12w^2$ on $y=0$. Hence
\eqref{eq:fo-strict-decay} gives $\dot{V}\le-m(a,A)<0$ almost everywhere
on this set, so sublevel sets of $V$ are forward invariant and $V$
falls below $a$ within a time depending on $(a,A)$ only. This is UGAS
of $(y,w)=0$ and, since $\combinedmapfo(0)=0$, of the stated error
origin; all bounds are uniform in the initial time and the admissible
disturbance for each fixed $b$.

2) Under \eqref{eq:fo-local-terminal}, the pair is homogeneous only asymptotically (cf. \cite{bhat2005geometric}), and integrating from $0$ yields
\begin{align*}
 \combinedmapfo(y)&=(\mu-\kappa)\spow{y}{\alpha}\bigl(1+o(1)\bigr),\\
 \potentialfo(y)&=\tfrac{1}{2}\kappa(\mu-\kappa)|y|^{2\alpha}\bigl(1+o(1)\bigr)
\end{align*}
as $y\to0$. Hence, near $y=0$, both $V$ and $w^2+\velprofilefo(y)^2$ are bounded above and below by positive
multiples of $w^2+|y|^{2\alpha}$, and $\combinedmapfo'\ge\tfrac{1}{2}\alpha(\mu-\kappa)|y|^{\alpha-1}$. Substitution in
\eqref{eq:fo-strict-decay}, using $|y|^{2\alpha}\le CV$ for some
$C>0$ and $\alpha<1$, gives
\begin{equation}
 \dot{V}\le-cV^\theta,\qquad
 c>0,\qquad
 \theta=\frac{3\alpha-1}{2\alpha}\in[1/2,1),
 \label{eq:fo-finite-time-decay}
\end{equation}
almost everywhere before settling, including $\alpha=1/2$ by the
crossing argument above.

Choose $\bar{v}>0$ sufficiently small that $V\le\bar{v}$ lies
within the region where these bounds hold. This
sublevel set is forward invariant, and uniform global attraction gives
finite entry time $t_{\rm e}$, uniformly on bounded sets of initial
errors. Integrating \eqref{eq:fo-finite-time-decay} then gives
\begin{equation*}
 T-t_{\rm e}\le
 \frac{V(y(t_{\rm e}),w(t_{\rm e}))^{1-\theta}}
      {c(1-\theta)}.
\end{equation*}
Hence the error origin is (uniformly) globally finite-time stable.
\end{proof}

Note that the proof uses the bound on $\dot{d}$ only along the
solution, so it suffices that
$|\dot{d}(t)|\le\delta\liminf_{r\to y(t)}|\velprofilefo(r)(\velprofilefo'(r)+\feedbackmapfo'(r))|$
for almost all $t$. For the PI example this yields
$|\dot{d}|\le\delta\kappa(\mu-\kappa)|y|$, i.e., the admissible rate
vanishes as $y\to0$, reflecting the well-known fact that only
constant disturbances are rejected exactly at the origin (and slowly
varying ones approximately). 
For super-twisting-like pairs (i.e., \eqref{eq:fo-local-terminal} with $\alpha=\tfrac{1}{2}$, as for the STA and its variants above), on
the other hand, it is $\delta\kappa(\mu-\kappa)/2$, including at
$y=0$.

Note also that
\eqref{eq:fo-dyn-controller} following from the pair
$(\velprofilefo,\feedbackmapfo)$ gives a useful freedom, as
taking simply
$\feedbackmapfo'=\injectionbound/|\velprofilefo|-\velprofilefo'$ with
$\feedbackmapfo(0)=0$, one gets
$\velprofilefo(\velprofilefo'+\feedbackmapfo')=-\injectionbound\sgn(y)$
in the dynamic part of \eqref{eq:fo-dyn-controller}. Hence one can
reshape $\velprofilefo$ while keeping the bound $\injectionbound>0$, as
long as $\velprofilefo$ keeps the square-root behaviour in
\eqref{eq:fo-local-terminal}, which keeps $\feedbackmapfo$ continuous
and gives $\mu=\kappa+2\injectionbound/\kappa$.

Moreover, since replacing $(\velprofilefo,\feedbackmapfo)$ by
$(\sqrt{L}\velprofilefo,\sqrt{L}\feedbackmapfo)$ multiplies
$\injectionbound$ by $L$,   any bounded rate can in fact be met by scaling the pair when $\injectionbound>0$ (a property we will utilize in
Section~\ref{sec:so-dyn}).

However, scaling the pair helps less when the rate bound is broken
outright, e.g., by an impulse, a step of a sampled input, or a load
that changes faster than a fixed gain allows. What such a break
costs is a delay set by its variation:
\begin{corollary}
\label{cor:fo-transient}
Let $(\velprofilefo,\feedbackmapfo)$ and $\delta$ be as in
Proposition~\ref{prop:first-order}, with \eqref{eq:fo-local-terminal}
holding for some $\alpha\in[1/2,1)$, and take $L\ge1$. Let $d=d_1+d_2$, where
$|\dot{d}_1|\le\delta L\injectionbound$ almost everywhere and
$d_2$ is constant outside an interval $[t_1,t_2]$ with
$\int_{t_1}^{t_2}|\dot{d}_2|\,dt\le\Delta_d$, where $d_2$ may jump
and $\dot{d}_2$ is its distributional derivative. Then, under
\eqref{eq:fo-dyn-controller} with
$(\sqrt{L}\velprofilefo,\sqrt{L}\feedbackmapfo)$, for every compact set of initial values of $(y,\xi+d/b)$ there are
$T,c>0$, depending on the set and $\Delta_d$, such
that
\begin{equation*}
  (y,\xi+d/b)=(0,0)\quad\text{for all }
 t\ge t_2+\frac{T}{\sqrt{L}}+c\,\frac{\Delta_d^{(1-\alpha)/\alpha}}{L^{1/(2\alpha)}}.
\end{equation*}
\end{corollary}
\begin{proof}
With $\xi=\sqrt{L}\tilde{\xi}$ and $\tau=\sqrt{L}t$, the loop with
$(\sqrt{L}\velprofilefo,\sqrt{L}\feedbackmapfo)$ is \eqref{eq:first-order} under
\eqref{eq:fo-dyn-controller} in time $\tau$ with disturbance $d/\sqrt{L}$,
whose rate bound is $\delta\injectionbound$ and whose break has
variation $\Delta_d/\sqrt{L}$; the initial errors stay in the compact set
for $L\ge1$. It thus suffices to take $L=1$. Split
$w\dot{d}/b=w\dot{d}_1/b+w\dot{d}_2/b$ in the proof of
Proposition~\ref{prop:first-order}. The first term is absorbed as
there, while the second gives $\dot{V}\le\sqrt{2V}\,|\dot{d}_2|/b$ on
$[t_1,t_2]$, so $\sqrt{V}$ grows by at most $\Delta_d/(b\sqrt2)$
there and decreases elsewhere. From $t_2$ on, the proof of
Proposition~\ref{prop:first-order} applies with
$\sqrt{V(t_2)}\le\sqrt{V(t_0)}+\Delta_d/(b\sqrt2)$, the settling time
being proportional to $(\sqrt{V})^{(1-\alpha)/\alpha}$ by
\eqref{eq:fo-finite-time-decay}. Thus, using $(r+s)^p\le r^p+s^p$
for any $r,s\ge0$ and $p\in(0,1]$, the growth of $\sqrt{V}$ by
$\Delta_d/(b\sqrt2)$ adds at most $c\,\Delta_d^{(1-\alpha)/\alpha}$
to the settling time counted from $t_2$. Returning to time $t$
divides all times by $\sqrt{L}$ and replaces $\Delta_d$ by $\Delta_d/\sqrt{L}$.
\end{proof}

The scaling gain $L$ thus shortens both the settling and the
recovery, but of course not the break itself.

\section{Dynamic feedback for second-order systems}\label{sec:so-dyn}
For the second-order system \eqref{eq:second-order}, the perturbation bound \eqref{eq:incremental-bound} forces
$d(t,q,0)=0$, so Theorem~\ref{thm:general-f} does not cover constant
loads such as gravity, friction offsets or actuator biases. We now remove this restriction by combining the two previous designs:
the static law \eqref{eq:general-controller} is retained, and the
integral action of Section~\ref{sec:fo} is used to estimate and cancel
the disturbance. First, however, we show how a simpler integral
extension can be added to reject constant disturbances.
\subsection{Integral action}

In Section~\ref{sec:fo}, integral action arose from correcting $\dot{u}$
in proportion to the mismatch between the actual and the prescribed
rate. The same step, one level up, corrects the input in proportion to
the mismatch between the actual and the prescribed acceleration:
\begin{equation*}
 \dot{u}=\dot{\nominalacc}+\lambda\bigl(\nominalacc-\ddot{q}\bigr),
\end{equation*}
with $\lambda>0$ and  $\nominalacc$ corresponding to the static law \eqref{eq:general-controller} (cf.~\eqref{eq:springdamper}):
\begin{equation}
 \nominalacc(q,\dot{q}):=\velenergyso'(q)+\feedbackmapso\bigl(q,\velprofileso(q)-\dot{q}\bigr)
 =-\restoringso(q)-\dampingso(q,\dot{q}).
 \label{eq:dob-static}
\end{equation}
Integrating yields a dynamic controller using only $q$
and $\dot{q}$:
\begin{subequations}
    \label{eq:so-dyn-controller}
\begin{align}
 u&=\nominalacc(q,\dot{q})-\lambda\dot{q}+\xi,\\
 \dot{\xi}&=\lambda\,\nominalacc(q,\dot{q}).
\end{align}
\end{subequations}
Defining $\hat{d}:=\lambda\dot{q}-\xi$, one then has $u=\nominalacc-\hat{d}$ and
$\dot{\hat{d}}=\lambda(\ddot{q}-u-\hat{d})$, so \eqref{eq:so-dyn-controller} is simply the static law combined with a first-order disturbance observer of
bandwidth $\lambda$ \cite{ohishi1987microprocessor,chen2016disturbance}.

For example, taking $\velprofileso(q)=-\kappa q$ and $\feedbackmapso(q,s)=\beta s$ with
$\beta>\kappa$, the controller is, up to the initial value of $\xi$, the
standard PID controller:
\begin{equation*}
 u=-k_Pq-k_D\dot{q}-k_I\!\int_0^t\! q\,d\tau,
\end{equation*}
with $k_P:=\kappa(\beta-\kappa)+\lambda\beta$, $k_D:=\beta+\lambda$ and
$k_I:=\lambda\kappa(\beta-\kappa)$. With $d=d_0-c\dot{q}$, the closed-loop
characteristic polynomial is $s^3+(bk_D+c)s^2+bk_Ps+bk_I$, which, by
the Routh--Hurwitz criterion, is Hurwitz if and only if
$(bk_D+c)k_P>k_I$. Since $k_Dk_P>k_I$, this holds for every $\lambda>0$ whenever $b\ge1$.

The next result extends this to general $\velprofileso$ and $\feedbackmapso$, showing that
$\lambda(b-1)+c\ge0$, in particular $b\ge1$, also suffices for global
asymptotic stability of the dynamic extension
\eqref{eq:so-dyn-controller} under constant perturbations:

\begin{proposition}
\label{prop:so-dyn}
Consider \eqref{eq:second-order} with $d=d_0-c\dot{q}$, where
$d_0\in\R$ and $c\ge0$ are constants, and let $\lambda>0$ satisfy
$\lambda(b-1)+c\ge0$. Let
 $\velprofileso$ and $\feedbackmapso$ satisfy
Defs.~\ref{def:velocity-profile} and~\ref{def:feedbackmap}. Then $(q,\dot{q},\xi)=(0,0,-d_0/b)$ is a globally asymptotically
stable equilibrium of the closed-loop system  \eqref{eq:second-order} under \eqref{eq:so-dyn-controller}.
\end{proposition}

\begin{proof}
With $\tilde{w}:=\xi+(d_0-\lambda\dot{q})/b$, and $\restoringso$, $\dampingso$,
$\potentialso$ as in the proof of Theorem~\ref{thm:general-f}, the closed loop is
\begin{equation*}
 \ddot{q}=-b\restoringso(q)-b\dampingso(q,\dot{q})+b\tilde{w}-bp\dot{q},\qquad
 \dot{\tilde{w}}=-\lambda\tilde{w}+\lambda p\dot{q},
\end{equation*}
where $p:=\lambda(1-b^{-1})+c/b\ge0$. If $p>0$, then
$V:=\tfrac12\dot{q}^2+b\potentialso(q)+\tfrac{b}{2\lambda p}\tilde{w}^2$ is positive
definite and radially unbounded, with
$\dot{V}=-b\dot{q}\dampingso(q,\dot{q})-b(\sqrt{p}\,\dot{q}-\tilde{w}/\sqrt{p})^2\le0$, and
$\dot{V}\equiv0$ implies $\dot{q}\equiv\tilde{w}\equiv0$ and thus $q\equiv0$;
the invariance principle \cite{lasalle1968stability,bacciotti1999stability} gives the result. If $p=0$, then $\{\tilde{w}=0\}$ is invariant and globally exponentially
stable, and solutions are bounded since $V_0:=\tfrac12\dot{q}^2+b\potentialso(q)$
satisfies $\dot{V}_0=-b\dot{q}\dampingso(q,\dot{q})+b\dot{q}\tilde{w}\le b\sqrt{2V_0}\,|\tilde{w}|$. On $\{\tilde{w}=0\}$, the
closed loop is that of Theorem~\ref{thm:general-f} with $d=0$,
so the result follows from Theorem~\ref{thm:general-f} and a
reduction principle \cite{el2013reduction,maggiore2018reduction}.
\end{proof}

\subsection{Exact rejection}

The integral action of Proposition~\ref{prop:so-dyn} removes the effect of constant
loads asymptotically; however, the prescribed motion is never
recovered exactly under time-varying loads in a finite time. 

To also reject time-varying disturbances, we will now  apply the dynamic feedback of Section~\ref{sec:fo} on top of the static feedback from Section~\ref{sec:so}.  To this end, we define the
error 
\begin{equation*}
    \velestimateerror:=\dot{q}-\velestimate
\end{equation*}
between $\dot{q}$ and the
velocity $\velestimate$ of the nominal motion. This is used in the proposed dynamic controller: 
\begin{subequations}
 \label{eq:so-fo-controller}
 \begin{align}
 u&=\nominalacc(q,\dot{q})-\feedbackmapfo(\velestimateerror)+\distestimate,\\
 \dot{\velestimate}&=\nominalacc(q,\dot{q}),\\
 \dot{\distestimate}&=\velprofilefo(\velestimateerror)\bigl(\velprofilefo'(\velestimateerror)+\feedbackmapfo'(\velestimateerror)\bigr).
 \end{align}
\end{subequations}
Here  $\nominalacc$ is defined in \eqref{eq:dob-static} and 
$(\velprofilefo,\feedbackmapfo)$ satisfies
Def.~\ref{def:first-order-profiles}. Since
$\dot{\velestimate}=\nominalacc$, one has
$\ddot{q}=\nominalacc(q,\dot{q})+\dot{\velestimateerror}$, so the
deviation from the nominal motion is precisely $\dot{\velestimateerror}$.
Moreover,
\begin{subequations}
    \begin{align}     
     \dot{\velestimateerror}&=b\bigl(-\feedbackmapfo(\velestimateerror)+\distestimate\bigr)+\tilde{d},
 \\
 \tilde{d}&:=(b-1)\nominalacc(q,\dot{q})+d,
 \label{eq:so-fo-error-d}
\end{align}
  \label{eq:so-fo-error}
\end{subequations}
so $\velestimateerror$ obeys \eqref{eq:first-order} under
\eqref{eq:fo-dyn-controller}, with $\xi=\distestimate$ and disturbance
$t\mapsto\tilde{d}(t,q(t),\dot{q}(t))$. In this sense,
\eqref{eq:so-fo-controller} acts as a disturbance observer
\cite{ohishi1987microprocessor,chen2016disturbance}, where $\distestimate$ converges to
$-\tilde{d}/b$, the equivalent control on $\{\velestimateerror=0\}$
\cite{utkin1996integral}, and thereby cancels the lumped disturbance. Hence, the
pair $(\velprofileso,\feedbackmapso)$ shapes the nominal motion, while
$(\velprofilefo,\feedbackmapfo)$ determines which disturbances are
rejected exactly.

If $b$ is known, scaling $u$ gives $b=1$, and thus $\tilde{d}=d$ in \eqref{eq:so-fo-error}, so
the following is a direct consequence of Theorem~\ref{thm:general-f}
and Proposition~\ref{prop:first-order}:
\begin{corollary}
\label{cor:so-fo}
Consider \eqref{eq:second-order} with $b=1$, where $d$ does not depend
on $(q,\dot{q})$, is locally absolutely continuous, and satisfies
$|\dot{d}|\le\delta\injectionbound$ almost everywhere, with
$\injectionbound$ as in \eqref{eq:fo-rate-bound} and some
$\delta\in[0,1)$. Let $(\velprofileso,\feedbackmapso)$ satisfy
Defs.~\ref{def:velocity-profile}--\ref{def:feedbackmap}, and
let $(\velprofilefo,\feedbackmapfo)$ satisfy
Def.~\ref{def:first-order-profiles} and
\eqref{eq:fo-local-terminal}. Then the origin of
$(q,\dot{q},\velestimateerror,\distestimate+d)$ is globally
asymptotically stable for the closed-loop system
\eqref{eq:second-order} under \eqref{eq:so-fo-controller}, and
globally finite-time stable if \eqref{eq:general-homogeneity} holds.
\end{corollary}

\begin{proof}
Since $\ddot{q}=\nominalacc(q,\dot{q})+\dot{\velestimateerror}$, take
$V:=\frac{1}{2}\dot{q}^2+\potentialso(q)$ with $\potentialso$ as in the
proof of Theorem~\ref{thm:general-f}. We have
$\dot{V}=-\dot{q}\dampingso(q,\dot{q})+\dot{q}\dot{\velestimateerror}
\le\sqrt{2V}\,|\dot{\velestimateerror}|$, so
\begin{equation}
 \sqrt{V(t)}\le\sqrt{V(0)}
 +\tfrac{1}{\sqrt2}\int_0^t|\dot{\velestimateerror}|\,ds.
 \label{eq:so-fo-energy}
\end{equation}
By Proposition~\ref{prop:first-order} with $\underline{b}=b=1$,
$(\velestimateerror,\distestimate+d)$ vanishes after a finite time $T$,
and $T$ and $\sup_t|\dot{\velestimateerror}|$ tend to zero with the
initial error. Hence \eqref{eq:so-fo-energy} rules out finite escape
and keeps $V(T)$ small for small initial data, and, for $t\ge T$,
Theorem~\ref{thm:general-f} applies with $b=1$ and $d=0$.
Under \eqref{eq:general-homogeneity}, it gives finite-time
convergence of $(q,\dot{q})$ from its state at $T$, which with the
finite-time convergence of $(\velestimateerror,\distestimate+d)$
gives that of the full error state.
\end{proof}

For $b\neq1$, the active disturbance $\tilde{d}$ in
\eqref{eq:so-fo-error-d} also contains the gain mismatch
$(b-1)\nominalacc$, whose rate grows with the state and does not vanish
with $\velestimateerror$. By Proposition~\ref{prop:first-order},
rejecting it therefore requires an integral action that does not vanish
at $\velestimateerror=0$, and that dominates this rate on the region of
interest. 
Taking inspiration from the super-twisting differentiator
\cite{levant1998robust} and high-gain observers
\cite{freidovich2008performance}, we achieve the latter by scaling
$(\velprofilefo,\feedbackmapfo)$ in \eqref{eq:so-fo-controller} by
$\sqrt{L}\,\injectionshape(q)$, with $L>0$ and
$\injectionshape(q)\ge1$, which by Section~\ref{sec:fo} multiplies
$\injectionbound$ by $L\injectionshape(q)^2$:
\begin{subequations}
 \label{eq:so-fo-controller-L}
 \begin{align}
 u&=\nominalacc(q,\dot{q})-\sqrt{L}\,\injectionshape(q)\feedbackmapfo(\velestimateerror)+\distestimate,\\
 \dot{\velestimate}&=\nominalacc(q,\dot{q}),\\
 \dot{\distestimate}&=L\,\injectionshape(q)^2\velprofilefo(\velestimateerror)\bigl(\velprofilefo'(\velestimateerror)+\feedbackmapfo'(\velestimateerror)\bigr).
 \end{align}
\end{subequations}
This scales the integral action by $L\injectionshape(q)^2$, so that a
nonconstant $\injectionshape$ shapes it with position. We refer to
\eqref{eq:so-fo-controller-L} as profile-anchored control with
exact rejection (PACER): the profile anchors the nominal motion to position,
and the first-order loop rejects the disturbance exactly. 
Note that, since the right-hand side of \eqref{eq:so-fo-controller-L}
may be discontinuous at $\velestimateerror=0$ (for $\alpha=\tfrac12$
in \eqref{eq:fo-local-terminal}), solutions are understood in the
Filippov sense, as in Section~\ref{sec:fo}.

However, the rate of $(b-1)\nominalacc$ must then be bounded on
compact sets, which requires a locally Lipschitz $\nominalacc$ and thus
excludes (terminal) profiles satisfying the homogeneity conditions  \eqref{eq:general-homogeneity} with $\alpha<1$. Thus finite-time convergence of $(q,\dot{q})$ is in general lost (we
return to this in Proposition~\ref{prop:so-fo-terminal}), but
the disturbance is still rejected exactly after a finite time, although only ensured semiglobally ($L$ must scale with the size of the initial conditions).
For the
same reason, $d$ must have a bounded rate in $t$ and be at least Lipschitz in
the state:
\begin{assumption}
\label{ass:perturbation}
The perturbation $d$ in \eqref{eq:second-order} is locally absolutely
continuous in $t$ and locally Lipschitz in $(q,\dot{q})$, uniformly in
$t$, with 
\begin{equation*}
 |d(t,0,0)|\le D_0,\qquad
 \Bigl|\tfrac{\partial d}{\partial t}(t,q,\dot{q})\Bigr|\le D_1,
\end{equation*}
holding for almost all $t$ and some
$D_0,D_1\ge0$.
\end{assumption}
Thus, if $\nominalacc$ is locally Lipschitz, then $\tilde{d}$ in \eqref{eq:so-fo-error-d} is  Lipschitz in
the state and has a bounded rate on every compact set.

We are now ready to state the main result of this paper:
\begin{theorem}[PACER]
\label{thm:so-fo}
Consider \eqref{eq:second-order}, i.e., $\ddot{q}=bu+d$, under the control
law \eqref{eq:so-fo-controller-L}, with $b\in[\underline{b},\bar{b}]$, for
$0<\underline{b}\le\bar{b}<\infty$ known, and $d$ satisfying
Assumption~\ref{ass:perturbation}. Let
\begin{enumerate}[label=\alph*), leftmargin=*]
 \item $(\velprofileso,\feedbackmapso)$ satisfy
 Defs.~\ref{def:velocity-profile}--\ref{def:feedbackmap}, with
 $\nominalacc$ in \eqref{eq:dob-static} locally Lipschitz;

 \item $(\velprofilefo,\feedbackmapfo)$ be as in Proposition~\ref{prop:first-order}, with $\alpha=\tfrac{1}{2}$ in \eqref{eq:fo-local-terminal}, $\Gamma>0$ in \eqref{eq:fo-rate-bound} and $\delta$ as in \eqref{eq:fo-rate-margin};
 \item $\injectionshape:\R\to[1,\infty)$ be locally Lipschitz.
\end{enumerate}
Then, for every compact set $\mathcal{K}\subset\R^4$ of initial
conditions $(q,\dot{q},\velestimate,\distestimate)(0)$, with $\tilde{d}$ as in
\eqref{eq:so-fo-error-d} and error state
$(q,\dot{q},\velestimateerror,\distestimate+\tilde{d}/b)$, there is an
$L^\star>0$, independent of $b$, such that for all $L\ge L^\star$ the
origin of the error state is  asymptotically stable with
region of attraction containing $\mathcal{K}$. Moreover, every
solution starting in $\mathcal{K}$ satisfies
$\ddot{q}=\nominalacc(q,\dot{q})$ after a finite time.
\end{theorem}
\begin{proof}
As in the proof of Corollary~\ref{cor:so-fo}, \eqref{eq:so-fo-energy}
holds. However, now the rate of $\tilde{d}$, which the loop
\eqref{eq:so-fo-error} must dominate, depends on the motion
$(q(t),\dot{q}(t))$, which in turn depends on this loop. Increasing $L$
makes the rate that the loop can dominate grow like $L$, whereas, as we
show, the rate of $\tilde{d}$ grows at most like $\sqrt{L}$. We
therefore i) rescale time so that the loop no longer depends on $L$,
ii) use this to confine $(q,\dot{q})$ to a compact set independent of
$L$, and iii) show that the rate of $\tilde{d}$ is dominated on this set
for $L$ large:

i) With $\rho:=\sqrt{L}\,\injectionshape(q)$, take
$d\tau/dt=\rho$,  and define
\begin{equation*}
 \tilde{\xi}:=\frac{\distestimate}{\rho}+\int_0^\tau\frac{\distestimate}{\rho}\,\frac{\dot{\rho}}{\rho^2}\,ds,\qquad
 D:=\frac{\tilde{d}}{\rho}-b\int_0^\tau\frac{\distestimate}{\rho}\,\frac{\dot{\rho}}{\rho^2}\,ds.
\end{equation*}
By using the notation $y'=\frac{d}{d\tau}y$, the following corresponds to \eqref{eq:so-fo-error} under \eqref{eq:so-fo-controller-L}:
\begin{equation*}
 \velestimateerror'=b\bigl(-\feedbackmapfo(\velestimateerror)+\tilde{\xi}\bigr)+D,\qquad
 \tilde{\xi}'=\velprofilefo(\velestimateerror)\bigl(\velprofilefo'(\velestimateerror)+\feedbackmapfo'(\velestimateerror)\bigr).
\end{equation*}
This is of the form \eqref{eq:first-order} under
\eqref{eq:fo-dyn-controller}, with $\xi=\tilde{\xi}$ and disturbance
$D$. Proposition~\ref{prop:first-order} therefore applies whenever
\begin{equation*}
 |D'|=\Bigl|\frac{\dot{\tilde{d}}}{\rho^2}-b\bigl(\tilde{\xi}+D/b\bigr)\frac{\dot{\rho}}{\rho^2}\Bigr|\le\delta\injectionbound,
\end{equation*}
with $\delta$ as in \eqref{eq:fo-rate-margin}.
Since $\tilde{\xi}+D/b=(\distestimate+\tilde{d}/b)/\rho$, with
$\tilde{d}$ bounded on compact sets by
Assumption~\ref{ass:perturbation}, and $\rho\ge\sqrt{L}$, the initial
errors lie, for all $L\ge1$, in a fixed compact sublevel set
$\mathcal{V}:=\{V\le\bar{V}\}$,
with $V$ as in \eqref{eq:fo-strict-function} using $w=\tilde{\xi}+D/b-\combinedmapfo(\velestimateerror)$, and with $\bar{V}$ independent of
$L$.

ii) If the condition on $D'$ holds,
Proposition~\ref{prop:first-order} ensures that the errors remain in
$\mathcal{V}$ and vanish before a finite time $\tau=T$, uniform on
$\mathcal{V}$. Moreover, by compactness, on $\mathcal{V}$,
\begin{equation*}
 |\velestimateerror'|\le Y,\qquad
 \Bigl|\frac{\distestimate+\tilde{d}/b}{\rho}\Bigr|\le E,
\end{equation*}
for some $Y,E>0$. Since $\mathcal{V}$ is independent of $L$, so are
$T$, $Y$ and $E$, and by continuity in $b$ they can be taken uniform
over $b\in[\underline{b},\bar{b}]$. Hence, while the rate condition
holds,
\begin{equation*}
 \int_0^t|\dot{\velestimateerror}(s)|\,ds
 =\int_0^{\tau(t)}|\velestimateerror'(s)|\,ds\le YT.
\end{equation*}
Thus, by \eqref{eq:so-fo-energy}, $(q,\dot{q})$ stays in a compact set
$\Omega$ independent of $L$. On smaller sublevel sets, $Y$ and $T$
tend to zero, so this integral tends to zero with the initial errors.

iii) By Assumption~\ref{ass:perturbation} and $\nominalacc$ locally
Lipschitz, $\tilde{d}=(b-1)\nominalacc+d$ (see
\eqref{eq:so-fo-error-d}) is Lipschitz in $(q,\dot{q})$ on $\Omega$
with some constant $\ell$, and its rate in $t$ is at most $D_1$.
Along solutions, almost everywhere,
$|\dot{\tilde{d}}|\le D_1+\ell(|\dot{q}|+|\ddot{q}|)$, where
$\ddot{q}=\nominalacc+\dot{\velestimateerror}$ with
$|\dot{\velestimateerror}|=\rho|\velestimateerror'|\le\rho Y$. Hence
$|\dot{\tilde{d}}|\le\ell(|\dot{q}|+|\nominalacc|+\rho Y)+D_1$. Thus,
with $\dot{\rho}=\sqrt{L}\,\injectionshape'(q)\dot{q}$,
$|\tilde{\xi}+D/b|\le E$ and $\rho\ge\sqrt{L}$,
\begin{align*}
 |D'|&\le\bigl(|\dot{\tilde{d}}|+\bar{b}\,|\tilde{\xi}+D/b|\,|\dot{\rho}|\bigr)\rho^{-2}\\
 &\le\bigl(\ell(|\dot{q}|+|\nominalacc|+\rho Y)+D_1+\bar{b}E\sqrt{L}\,|\injectionshape'|\,|\dot{q}|\bigr)\rho^{-2}\\
 &\le \left({\ell(|\dot{q}|+|\nominalacc|)+D_1}\right)/{L}+\left({\ell Y+\bar{b}E\,|\injectionshape'|\,|\dot{q}|}\right)/{\sqrt{L}}.
\end{align*}
Since $\nominalacc$, $\dot{q}$ and $\injectionshape'$ are bounded on
$\Omega$, the right-hand side is at most $c_1/L+c_2/\sqrt{L}$ on
$\Omega\times\mathcal{V}$, for constants $c_1,c_2$ independent of $L$
and, by ii), of $b$. Hence it is at most $\delta\injectionbound$ for
all $L\ge L^\star$, with $L^\star$ the solution of
$c_1/L^\star+c_2/\sqrt{L^\star}=\delta\injectionbound$.

Thus, for any compact set $\mathcal{K}\subset\R^4$ containing the
origin there is an $L^\star$ as in iii), and taking $L\ge L^\star$,
the condition on $D'$ holds for all $t$, so the errors vanish in finite time by
Proposition~\ref{prop:first-order}. After this,
$\ddot{q}=\nominalacc(q,\dot{q})$ for any $b$, and asymptotic
stability, with a region of attraction containing $\mathcal{K}$,
follows as in Corollary~\ref{cor:so-fo}.
\end{proof}

\begin{remark}
A practical starting value when choosing $L$ can be
\begin{equation}
    \label{eq:L-rule-of-thumb}
 L^\star\approx\frac{\sup|\dot{\tilde{d}}|}{\delta\injectionbound}.
\end{equation}
Indeed, the proof gives $L^\star$ explicitly, but conservatively, and
with $\velestimate(0)=\dot{q}(0)$ and $\distestimate(0)=0$ it reduces
to the above. Since $\dot{\tilde{d}}=(b-1)\dot{\nominalacc}+\dot{d}$, with $\sup|\dot{\nominalacc}|$ along the
nominal motion, $b\ge1$ after normalization, and an estimate of
$\sup|\dot{d}|$, take
$L\approx\bigl((\bar{b}-1)\sup|\dot{\nominalacc}|+\sup|\dot{d}|\bigr)/(\delta\injectionbound)$,
and increase from there.
\end{remark}

Note that Theorem~\ref{thm:so-fo} requires $\injectionbound>0$, since
near $\velestimateerror=0$ the rejection loop in
\eqref{eq:so-fo-controller-L} must behave as a super-twisting
algorithm, as the mismatch rate $D'$ in the proof does not vanish there. Away from $\velestimateerror=0$ the pair can be shaped
differently, within Def.~\ref{def:first-order-profiles}, and the gain
is scheduled by $L\injectionshape(q)^2$. The super-twisting pair
$\velprofilefo(\velestimateerror)=-\kappa\spow{\velestimateerror}{1/2}$,
$\feedbackmapfo(\velestimateerror)=\mu\spow{\velestimateerror}{1/2}$,
$\mu>\kappa>0$, gives $\injectionbound=\kappa(\mu-\kappa)/2$ in
\eqref{eq:fo-rate-bound}. PI, and any pair with $\alpha>\frac12$ in
\eqref{eq:fo-local-terminal}, gives $\injectionbound=0$ and is not
covered as its integral action stalls as $\velestimateerror\to0$. A global version would require $\injectionshape$ to depend at least
also on $\velestimate$, with growth bounds on $\nominalacc$ and $d$ as
in variable-gain super-twisting control \cite{gonzalez2012variable};
we do not pursue it here.

Note also that, while Theorem~\ref{thm:so-fo} restores the nominal loop
exactly after a finite time, it says nothing about what happens during
this transient. In particular, the velocity envelope of
Corollary~\ref{cor:envelope} may be violated, as the injection
$-\sqrt{L}\,\injectionshape\feedbackmapfo(\velestimateerror)+\distestimate$
pushes $\dot{q}$ towards the nominal velocity $\velestimate$, not
towards the profile, so neither $|\dot{q}|\le|\velprofileso(q)|$ nor
that $q$ keeps its sign is guaranteed before the errors vanish. The
next result shows that this transient is in fact harmless. If
$\feedbackmapso$ acts in the direction of the deviation, both hold up
to a margin that vanishes as $L$ grows.
\begin{corollary}
\label{cor:pace-envelope}
Let the hypotheses of Theorem~\ref{thm:so-fo} hold with
$s\feedbackmapso(q,s)\ge0$ for all $(q,s)\in\R^2$, and take
$\velestimate(0)=\dot{q}(0)$, $\distestimate(0)=0$. Then, for every
compact set $\mathcal{K}_0$ of initial states $(q(0),\dot{q}(0))$ with
$|\dot{q}(0)|\le|\velprofileso(q(0))|$, there are $L^\star>0$ and
$\chi(L)=O(L^{-1/2})$, both independent of $b$, such that for
$L\ge L^\star$ every solution starting in $\mathcal{K}_0$ satisfies
\begin{equation*}
 |\dot{q}(t)|\le|\velprofileso(q(t))|+\chi(L)\quad\text{for all } t\ge0.
\end{equation*}
Moreover, whenever $q(t)q(0)<0$,
\begin{equation*}
 \int_{q(t)}^{0}\nominalacc(r,0)\,dr\le\chi(L)^2,
\end{equation*}
which bounds any overshoot of the target $q=0$.
\end{corollary}
\begin{proof}
Let $W:=\tfrac12\dot{q}^2-\velenergyso(q)$, so that the velocity bound
reads $W\le\chi^2/2$. Since
$\ddot{q}=\nominalacc+\dot{\velestimateerror}$ and
$\nominalacc=\velenergyso'+\feedbackmapso(q,\velerrorso)$,
$\dot{W}=\dot{q}\,\feedbackmapso(q,\velerrorso)+\dot{q}\,\dot{\velestimateerror}$.
Where $W>0$, $\velerrorso$ has the sign of $-\dot{q}$, so
$s\feedbackmapso(q,s)\ge0$ gives
$\dot{W}\le|\dot{q}||\dot{\velestimateerror}|$. Take $L\ge L^\star$
from Theorem~\ref{thm:so-fo} for the initial set
$\{(q,\dot{q},\dot{q},0):(q,\dot{q})\in\mathcal{K}_0\}$. By ii) of its
proof, $(q,\dot{q})$ stays in $\Omega$ and
$\int_0^\infty|\dot{\velestimateerror}|\,dt\le YT$. Integrating
$\dot{W}$ from the last time $W\le0$, which exists since $W(0)\le0$,
gives $W\le YT\max_{\Omega}|\dot{q}|=:\chi(L)^2/2$.

For the overshoot, $V_0=\tfrac12\dot{q}^2+\potentialso(q)$ from the
proof of Theorem~\ref{thm:general-f}, with $b=1$, satisfies
$\dot{V}_0=-\dot{q}\dampingso(q,\dot{q})+\dot{q}\,\dot{\velestimateerror}
\le|\dot{q}||\dot{\velestimateerror}|$, and
$\potentialso(q)=\int_q^0\nominalacc(r,0)\,dr$ by
\eqref{eq:phipsi-def}. At the first time $t_c$ with $q=0$,
$V_0(t_c)=\tfrac12\dot{q}(t_c)^2\le\chi^2/2$ by the velocity bound, so
for $t\ge t_c$,
$\potentialso(q(t))\le V_0(t)\le\chi^2/2+YT\max_{\Omega}|\dot{q}|=\chi^2$.

Finally, by i) of the proof of Theorem~\ref{thm:so-fo}, the initial
errors are $(0,\tilde{d}(0)/(b\rho(0)))$, with $\tilde{d}(0)$ bounded
on $\mathcal{K}_0$ and $\rho(0)\ge\sqrt{L}$, so they lie in a sublevel
set of \eqref{eq:fo-strict-function} whose level is $O(1/L)$. By ii),
$Y$ and $T$ tend to zero with the level, uniformly in $b$, so
$\chi(L)\to0$; and by \eqref{eq:fo-local-terminal} with
$\alpha=\tfrac12$, \eqref{eq:fo-finite-time-decay} with
$\theta=\tfrac12$ gives $Y,T=O(L^{-1/2})$, hence
$\chi(L)=O(L^{-1/2})$.
\end{proof}

Theorem~\ref{thm:so-fo} and Corollary~\ref{cor:pace-envelope} exclude
profiles with finite-time convergence, i.e.,
\eqref{eq:general-homogeneity} with $\alpha<1$. This is because
$\nominalacc$ is then not Lipschitz on $\{q=0\}\cup\{\velerrorso=0\}$,
so that the gain mismatch $(b-1)\nominalacc$ in $\tilde{d}$ has an
unbounded rate there, which the rejection loop cannot dominate. The
following provides additional conditions under which the disturbance
is  rejected exactly and the target reached in finite
time:
\begin{proposition}
\label{prop:so-fo-terminal}
Let the hypotheses of Theorem~\ref{thm:so-fo} hold with
$s\feedbackmapso(q,s)\ge0$ for all $(q,s)\in\R^2$, but with
$\nominalacc$ in \eqref{eq:dob-static} locally Lipschitz away from
$\{q=0\}\cup\{\velerrorso=0\}$, and with\footnote{For the sake of space, we consider the specific pair \eqref{eq:hom_pair} rather than a general homogeneous pair satisfying \eqref{eq:general-homogeneity} (or, for that matter, one that is only asymptotically homogeneous; see, e.g., \cite{bhat2005geometric}).}
\begin{equation}\label{eq:hom_pair}
 \velprofileso(q)=-\kappa\spow{q}{\alpha},\qquad
 \feedbackmapso(q,s)=\beta\spow{s}{\frac{2\alpha-1}{\alpha}},
\end{equation}
in a neighborhood of $q=0$ and of $\{s=0\}$, respectively, with
$\alpha\in[\tfrac{2}{3},1)$ and $\beta>\alpha\kappa^{1/\alpha}$.
Further, suppose $\mathcal{K}_0$ is a compact set of initial states
$(q(0),\dot{q}(0))$ with $|\dot{q}(0)|\le|\velprofileso(q(0))|$, and
that one of the following holds on all of $\mathcal{K}_0$:
\begin{enumerate}[label=\alph*)]
 \item $\distestimate(0)=0$ and $q(0)\neq0$;
 \item $\distestimate(0)=-\tilde{d}(0,q(0),\dot{q}(0))/b$.
\end{enumerate}
Then, if $\velestimate(0)=\dot{q}(0)$, there is an $L^\star>0$,
independent of $b$, such that for $L\ge L^\star$ every solution
starting in $\mathcal{K}_0$ satisfies
\begin{enumerate}[label=\roman*)]
 \item $(\velestimateerror,\distestimate+\tilde{d}/b)=(0,0)$ after a
 finite time, and for all $t\ge0$ under b), with $\tilde{d}$ as in
 \eqref{eq:so-fo-error-d};
 \item $|\dot{q}(t)|\le|\velprofileso(q(t))|+\chi(L)$ for all $t\ge0$,
 with $\chi(L)=O(L^{-1/2})$ as in Corollary~\ref{cor:pace-envelope},
 and $\chi=0$ under b);
 \item $(q,\dot{q})=(0,0)$ within a finite time.
\end{enumerate}
\end{proposition}
\begin{proof}
Let $\gamma:=\tfrac{2\alpha-1}{\alpha}\in[\tfrac{1}{2},1)$ and
$W:=\tfrac{1}{2}\dot{q}^2-\velenergyso(q)$ as in
Corollary~\ref{cor:pace-envelope}. All constants below are independent
of $L$ and of $b\in[\underline{b},\bar{b}]$. We proceed in five steps,
by showing that: 1) the errors vanish within a short time $T_1$; 2) $W\le\chi(L)^2/2$ on $[0,T_1]$; 3) exact rejection then persists; 4) $W\le0$ holds
before the target; 5) Theorem~\ref{thm:general-f} gives finite-time
arrival.

1) We first show that the errors vanish within a time $T_1=o(1)$ with
$\int_0^{T_1}|\dot{\velestimateerror}|\,dt=O(1/L)$ and $\ddot{q}$
bounded: under b) one has $T_1=0$, while under a),
$\velestimateerror(0)=\distestimate(0)=0$ by
$\velestimate(0)=\dot{q}(0)$, and $q(0)$ is bounded away from zero on
$\mathcal{K}_0$. Split $\tilde{d}$ into its Lipschitz part
$a:=(b-1)\velenergyso'(q)+d$ and $(b-1)\feedbackmapso(q,\velerrorso)$,
then define $z:={(\distestimate+a/b)}/{\rho}$ and $p:=\velprofileso(q)-\velestimate=\velerrorso+\velestimateerror$
with $\rho:=\sqrt{L}\injectionshape(q)$.
Taking $d\tau=\rho\,dt$ and $y':=\tfrac{d}{d\tau}y$, then
\begin{align*}
 \velestimateerror'&=b\big(z-\feedbackmapfo(\velestimateerror)\big)+(b-1)\feedbackmapso(q,\velerrorso)/\rho,\\
 z'&=\velprofilefo(\velestimateerror)\big(\velprofilefo'(\velestimateerror)+\feedbackmapfo'(\velestimateerror)\big)+e,\quad
 p'=-\big(\velprofileso'(q)\velerrorso+\feedbackmapso(q,\velerrorso)\big)/\rho,
\end{align*}
follow from \eqref{eq:so-fo-controller-L},
with $|e|=O\big((1+|z|+|\feedbackmapfo(\velestimateerror)|)/\sqrt{L}\big)$
by Assumption~\ref{ass:perturbation}, and $|z(0)|=O(1/\sqrt{L})$.

If $|\velerrorso(0)|\ge c_0/L$, then, by
\eqref{eq:hom_pair}, $\nominalacc$ is Lipschitz with
constant $O(L^{1-\gamma})$ on $\{|\velerrorso|\ge c_0/(2L)\}$. Therein, the proof
of Theorem~\ref{thm:so-fo} applies, with its scaled disturbance rate
$O(L^{-\gamma}+L^{-1/2})=o(1)$ by its step iii), and since the initial
level in \eqref{eq:fo-strict-function} is $O(1/L)$, the errors vanish
within $T_1=O(1/L)$, with
$\int_0^{T_1}|\dot{\velestimateerror}|\,dt=O(1/L)$ and $\ddot{q}$
bounded, as in the proof of Corollary~\ref{cor:pace-envelope}. In this
time $\velerrorso$ moves by $O(1/L)$, so $|\velerrorso|\ge c_0/(2L)$
indeed holds up to $T_1$ for $c_0$ sufficiently large.

If instead $|\velerrorso(0)|<c_0/L$, note that by
\eqref{eq:fo-local-terminal} with $\alpha=\tfrac{1}{2}$,
$\velprofilefo(\velprofilefo'+\feedbackmapfo')=-\tfrac{1}{2}\kappa(\mu-\kappa)\operatorname{sgn}(\velestimateerror)+o(1)$
and $\feedbackmapfo(\velestimateerror)=\mu\spow{\velestimateerror}{1/2}+o(|\velestimateerror|^{1/2})$
near $\velestimateerror=0$. For some $C>0$ to be chosen, let
\begin{equation*}
 V:=\frac{1}{2}z^2+\frac{\kappa(\mu-\kappa)}{2b}|\velestimateerror|+C|p|-\varepsilon\frac{z\velestimateerror}{\sqrt{z^2+|\velestimateerror|}},
\end{equation*}
the last term being zero at $\velestimateerror=z=0$, with
$\varepsilon>0$ so small that $V$ is bounded above and below by
positive multiples of $z^2+|\velestimateerror|+|p|$. For
$|\velestimateerror|$, $|\velerrorso|$ and $1/L$ small, one has
\begin{align*}
 V'\le&-\tfrac{1}{2}\kappa\mu(\mu-\kappa)|\velestimateerror|^{1/2}
   -\frac{\varepsilon b}{2}\frac{z^2}{\sqrt{z^2+|\velestimateerror|}}
   \\
   &+O\big(\varepsilon|\velestimateerror|^{1/2}+|z||e|+|\velerrorso|^\gamma/\rho\big)
   +C\operatorname{sgn}(p)p'\\
  \le&-c\big(|z|+|\velestimateerror|^{1/2}\big)+O(|\velerrorso|^\gamma/\rho)+C\operatorname{sgn}(p)p'.
\end{align*}
Here $|\velprofileso'\velerrorso|\le(\beta/2)|\velerrorso|^\gamma$ for
small $\velerrorso$, so $C\operatorname{sgn}(p)p'\le-C\beta|\velerrorso|^\gamma/(2\rho)$
when $p\velerrorso>0$, which for $C$ sufficiently large dominates the
$O(|\velerrorso|^\gamma/\rho)$ term, that of $(b-1)\feedbackmapso/\rho$ in
$\velestimateerror'$, while for $p\velerrorso\le0$ both are
$O(|\velestimateerror|^{1/2}/\rho)$, as $|\velerrorso|\le|\velestimateerror|$
there, and are absorbed by the first term for $L$ sufficiently large. On sliding
segments $\velestimateerror\equiv0$, where $p=\velerrorso$ and
$|z|=|b-1|\beta|\velerrorso|^\gamma/(b\rho)$, the decay of $C|p|$
dominates $zz'$ for $C$ sufficiently large as well. Hence, the following holds
almost everywhere along Filippov solutions:
\begin{equation*}
 \dot{V}\le-c\big(\rho|z|+\rho|\velestimateerror|^{1/2}+|\velerrorso|^\gamma\big)\le-cV^\gamma.
\end{equation*}
Moreover, $|\dot{\velestimateerror}|\le-C\dot{V}$, so
$\int_0^{T_1}|\dot{\velestimateerror}|\,dt\le CV(0)$ and
$\dot{\velestimateerror}=O(1)$. As
$V(0)=\tfrac{1}{2}z(0)^2+C|\velerrorso(0)|=O(1/L)$, this gives
$T_1=O(L^{-(1-\gamma)})$ and
$\int_0^{T_1}|\dot{\velestimateerror}|\,dt=O(1/L)$.

2) To show that $W\le\chi(L)^2/2$ on $[0,T_1]$ with
$\chi(L)=O(L^{-1/2})$, note that $\dot{q}\,\feedbackmapso(q,\velerrorso)\le0$
whenever $W>0$ and that $\dot{q}$ is bounded on $[0,T_1]$ by 1). Hence,
for $W>0$,
\begin{equation*}
 \dot{W}=\dot{q}\,\feedbackmapso(q,\velerrorso)+\dot{q}\,\dot{\velestimateerror}
 \le C|\dot{\velestimateerror}|.
\end{equation*}
Since $W(0)\le0$, 1) gives  $W\le C\int_0^{T_1}|\dot{\velestimateerror}|\,dt\le C/L$
on $[0,T_1]$.

3) We now show that exact rejection persists on
$\{|q|\ge r/2\}\cup\{W\le0\}$ for any fixed $r>0$ and $L$ sufficiently large: near
$\velerrorso=0$, by \eqref{eq:hom_pair} and
$\dot{\velerrorso}=-\velprofileso'\velerrorso-\beta\spow{\velerrorso}{\gamma}-\dot{\velestimateerror}$,
one has
\begin{equation*}
 \dot{\nominalacc}=\velenergyso''\dot{q}
 -\beta\gamma|\velerrorso|^{\gamma-1}\big(\velprofileso'\velerrorso+\beta\spow{\velerrorso}{\gamma}\big)
 -\beta\gamma|\velerrorso|^{\gamma-1}\dot{\velestimateerror}.
\end{equation*}
The first two terms are bounded on $\{|q|\ge r/2\}$, and on
$\{W\le0\}$ near $q=0$, where
$|\dot{q}|,|\velerrorso|\le2|\velprofileso(q)|$ makes them
$O(|q|^{3\alpha-2})$. Since $\dot{\velestimateerror}=0$ on the nominal
loop, $\dot{\tilde{d}}=(b-1)\dot{\nominalacc}+\dot{d}$ is bounded there
by Assumption~\ref{ass:perturbation}. Off it, the last term is at most
$\beta\gamma b|\velerrorso|^{\gamma-1}\big(|\distestimate+\tilde{d}/b|+\rho|\feedbackmapfo(\velestimateerror)|\big)$,
as $\dot{\velestimateerror}=b(\distestimate+\tilde{d}/b)-b\rho\feedbackmapfo(\velestimateerror)$.
Hence, for $\velerrorso\neq0$, the rate bound of
Proposition~\ref{prop:first-order} still holds for $L$ sufficiently large on the
neighborhood
$\{|\distestimate+\tilde{d}/b|+\rho|\velestimateerror|^{1/2}\le c|\velerrorso|^{1-\gamma}\}$
of the error origin, so \eqref{eq:fo-strict-function} is nonincreasing
there and zero errors stay zero.

For $\velerrorso=0$ and $q\neq0$, one has $V=0$, and the bound
$\dot{V}\le-cV^\gamma$ of 1), which required of $q$ only that
$|\velprofileso'\velerrorso|\le(\beta/2)|\velerrorso|^\gamma$, holds
on a neighborhood of $\velerrorso=0$; thus $V\equiv0$ persists.

At the origin, let $d_{\rm o}:=d(t,0,0)$,
$z_{\rm o}:=(\distestimate+d_{\rm o}/b)/\rho$ and $Q:=V-C|p|$ with
$z_{\rm o}$ in place of $z$. Then
$\velestimateerror'=b(z_{\rm o}-\feedbackmapfo(\velestimateerror))+(\tilde{d}-d_{\rm o})/\rho$
and $z_{\rm o}'=\velprofilefo(\velprofilefo'+\feedbackmapfo')+e$ with
$e=\dot{d}_{\rm o}/(b\rho^2)-z_{\rm o}\dot{\rho}/\rho^2=O(1/\sqrt{L})$,
so the bound on $V'$ in 1), with $C=0$ and $(b-1)\feedbackmapso/\rho$
replaced by $(\tilde{d}-d_{\rm o})/\rho$, gives
$\dot{Q}\le-c\rho\sqrt{Q}+C|\tilde{d}-d_{\rm o}|$.

The nominal Lyapunov function $V_\varepsilon(q,\velestimate)$ of the
proof of Theorem~\ref{thm:general-f} satisfies
$\dot{V}_\varepsilon\le-a_0V_\varepsilon^\theta$ near the origin, with
$a_0>0$ and $\theta:=(3\alpha-1)/(2\alpha)$, and by homogeneity
\begin{equation*}
    |\partial V_\varepsilon/\partial q|\le CV_\varepsilon^{\gamma/2} \quad \text{and} \quad 
|\partial V_\varepsilon/\partial\velestimate|\le CV_\varepsilon^{1/2}.
\end{equation*}
Since $\dot{q}=\velestimate+\velestimateerror$, and $\nominalacc$ by
\eqref{eq:hom_pair} satisfies
$|\nominalacc(q,\velestimate+\velestimateerror)-\nominalacc(q,\velestimate)|\le2\beta|\velestimateerror|^\gamma$,
Young's inequality gives
\begin{equation*}
 \dot{V}_\varepsilon\le-\tfrac{1}{2}a_0V_\varepsilon^\theta+CQ^{1+\gamma},
\end{equation*}
as $2\theta=1+\gamma$. Moreover,
$|\tilde{d}-d_{\rm o}|\le C\big(V_\varepsilon^{\gamma/2}+Q^\gamma\big)$ by
\eqref{eq:hom_pair} and Assumption~\ref{ass:perturbation}. Hence $U:=V_\varepsilon+Q^2$ satisfies
\begin{align*}
 \dot{U}\le &-\tfrac{1}{2}a_0V_\varepsilon^\theta-2c\rho Q^{3/2}
            +CQV_\varepsilon^{\gamma/2}+CQ^{1+\gamma}
 \\
        \le &-\tfrac{1}{4}a_0V_\varepsilon^\theta-c\rho Q^{3/2}\le0
\end{align*}
near the origin, by Young's inequality and
$CQ^{1+\gamma}\le c\rho Q^{3/2}$ for $L$ sufficiently large, as
$1+\gamma\ge\tfrac{3}{2}$. Thus, the origin, where $U=0$, is invariant
for every Filippov solution. 

Finally, if $\dot{\velestimateerror}=0$, one has
$\dot{W}=\dot{q}\,\feedbackmapso(q,\velerrorso)\le0$ on $\{W>0\}$ by 2),
 and $W(T_1)\le\chi^2/2$
gives $W\le\chi^2/2$ for all $t\ge T_1$.

4) We now show that $W\le0$ holds before the target. Under b),
$W(0)\le0$ already. Under a), take
$r<\tfrac{1}{2}\min_{\mathcal{K}_0}|q(0)|$ so small that
\eqref{eq:hom_pair} holds on $\{|q|\le r,\,W\le\chi^2/2\}$ for $L$
sufficiently large. The nominal loop reaches $\{|q|\le r\}$ by
Theorem~\ref{thm:general-f}, and if $W>0$ there, then
$|\dot{q}|\ge|\velprofileso(r/2)|$ and $W=O(|\velerrorso|)$ while
$|q|\ge r/2$, so $\dot{W}=-\beta|\dot{q}||\velerrorso|^\gamma\le-cW^\gamma$.
As $W=O(1/L)$ by 2), $W(T_2)=0$ at some $T_2$ within
$O(L^{-(1-\gamma)})$, with $|q(T_2)|>r/2$ for $L$ sufficiently large.

5) Finally, $W(T_2)\le0$, with $T_2=0$ under b), gives $W\le0$ for all
$t\ge T_2$ by 3), and Theorem~\ref{thm:general-f} yields
$(q,\dot{q})=(0,0)$ within a finite time.
\end{proof}

Note that, unlike Corollary~\ref{cor:envelope},
Corollary~\ref{cor:pace-envelope} and
Proposition~\ref{prop:so-fo-terminal} place no condition on $b$, $d$
or the monotonicity of the profile beyond those of
Theorem~\ref{thm:so-fo}, only on the sign of the feedback map. Indeed,
the gain mismatch and the perturbation are now absorbed by the
first-order loop. The velocity envelope and the absence of overshoot
are thus kept up to margins set by the injection gain and, for the finite-time designs, exactly after a finite time
(Proposition~\ref{prop:so-fo-terminal}). This is a key trait of PACER: the profile can be replanned and
switched safely mid-motion.

\section{Profile replanning via convex optimization}\label{sec:replanning}

A velocity profile is a curve in the $(q,\dot{q})$-plane, and the
controller only asks how far the state is from it. Replacing the
profile along the way (because of, say, a new target, a changed speed
restriction, an obstacle, etc.) therefore simply means drawing a new
curve through the current state. PACER is made for this: if the new curve passes through the current
state and matches the current acceleration, the rejection loop keeps
cancelling the disturbance across the switch, so the new profile is
followed exactly from the first instant and the switch is invisible
in the input.

Indeed, suppose that PACER runs with the triple
$(\velprofileso_1,\feedbackmapso_1,\injectionshape_1)$ and that, at
some time $t_s$, it is to be replaced by
$(\velprofileso_2,\feedbackmapso_2,\injectionshape_2)$ without
disturbing the motion. Two conditions make the switch invisible in the
input: 1) $\velprofileso_2$ passes through the current state, so that
$\velerrorso=0$ and the feedback map is at zero; 2) the nominal
acceleration of $\velprofileso_2$ at the current position equals the
one being applied. Then $\nominalacc$, and with it
$\dot{\velestimate}$ and $\tilde{d}=(b-1)\nominalacc+d$ in
\eqref{eq:so-fo-error-d}, are continuous at $t_s$, so the rejection
loop sees nothing and its estimate carries over. Since the target may
change at $t_s$, we write $q(t_s^-)$ and $q(t_s^+)$ for the position
at $t_s$ measured from the old and the new target, respectively:
\begin{proposition}
\label{prop:replanning}
For $i\in\{1,2\}$, let Theorem~\ref{thm:so-fo} apply with
$(\velprofileso_i,\feedbackmapso_i,\injectionshape_i)$. Let
$\nominalacc_i$ and $\tilde{d}_i$ be given by \eqref{eq:dob-static}
and \eqref{eq:so-fo-error-d} for this triple, let $q$ be measured
from the target of the active profile, and denote
$e_i:=(\velestimateerror,\distestimate+\tilde{d}_i/b)$. Suppose
$(\velprofileso_1,\feedbackmapso_1,\injectionshape_1)$ is switched to
$(\velprofileso_2,\feedbackmapso_2,\injectionshape_2)$ at a time
$t_s$ at which
\begin{subequations}
 \label{eq:splice-both}
 \begin{align}
  \velprofileso_2(q(t_s^+))&=\dot{q}(t_s^-),
  \label{eq:splice}\\
  \velenergyso_2'(q(t_s^+))&=\nominalacc_1(q(t_s^-),\dot{q}(t_s^-)),
  \label{eq:splice-acc}
 \end{align}
\end{subequations}
and either $\injectionshape_2(q(t_s^+))=\injectionshape_1(q(t_s^-))$
or $e_1(t_s^-)=0$, where $\velestimate$ and $\distestimate$ are
carried over, with only $\dot{\velestimate}=\nominalacc_i$ changing
at $t_s$. Then, for every compact set $\mathcal{K}\subset\R^4$ there
are $L^\star>0$ and $T\ge t_s$, independent of $b$, such that for
$L\ge L^\star$ every solution starting in $\mathcal{K}$ at some
$t_0<t_s$ satisfies:
\begin{enumerate}[label=\roman*), leftmargin=*]
 \item $\nominalacc$, $u$ and $\ddot{q}$ are continuous at $t_s$, and
 $e_2(t_s^+)=e_1(t_s^-)$;
 \item $e_2(t)=0$ and $\ddot{q}=\nominalacc_2(q,\dot{q})$ for all
 $t\ge T$;
 \item if $e_1(t_s^-)=0$, then $\dot{q}=\velprofileso_2(q)$ for all
 $t\ge t_s$.
\end{enumerate}
\end{proposition}
\begin{proof}
i) By \eqref{eq:splice}, $\velerrorso=0$ for $\velprofileso_2$ at
$t_s^+$, so
$\nominalacc_2(q(t_s^+),\dot{q}(t_s^-))=\velenergyso_2'(q(t_s^+))$
by \emph{ii} of Def.~\ref{def:feedbackmap}, and \eqref{eq:splice-acc}
makes $t\mapsto\nominalacc(q(t),\dot{q}(t))$ continuous at $t_s$.
Hence so are $\dot{\velestimate}$ and $\tilde{d}$ in
\eqref{eq:so-fo-error-d}, while $\velestimateerror$ and
$\distestimate$ are continuous, as $\velestimate$ and
$\distestimate$ are carried over; thus $e_2(t_s^+)=e_1(t_s^-)$. With
$\distestimate$ carried over and $\injectionshape$ continuous at
$t_s$, or $\velestimateerror(t_s)=0$, the injection is continuous as
well, hence so are $u$ and $\ddot{q}$.

ii) Let $L_1^\star$ be given by Theorem~\ref{thm:so-fo} for
$(\velprofileso_1,\feedbackmapso_1,\injectionshape_1)$ and
$\mathcal{K}$. For $L\ge L_1^\star$, by ii) of its proof,
$(q,\dot{q})$ stays in a compact set $\Omega_1$ and the rescaled
errors in $\mathcal{V}$ up to $t_s$, both independent of $L$. By i),
the rescaled errors are still in $\mathcal{V}$ at $t_s^+$, since
$(\velprofilefo,\feedbackmapfo)$ is unchanged and either $\rho$ is
continuous at $t_s$ or the errors are zero; moreover,
\eqref{eq:so-fo-energy} for $\velprofileso_2$ starts from an energy
bounded on $\Omega_1$. Hence the proof of Theorem~\ref{thm:so-fo}
applies to $(\velprofileso_2,\feedbackmapso_2,\injectionshape_2)$
from $t_s$, which gives $L_2^\star$, and, by its step ii), the errors
vanish before a $\tau$-time $T_\tau$ after the switch, uniform on
$\mathcal{V}$ and independent of $b$. With
$L^\star:=\max\{L_1^\star,L_2^\star\}$ and $\rho\ge\sqrt{L}\ge1$,
this is before $t_s+T_\tau=:T$.

iii) If $e_1(t_s^-)=0$, then $e_2(t_s^+)=0$ by i), and
Proposition~\ref{prop:first-order} keeps it there, so
$\ddot{q}=\nominalacc_2(q,\dot{q})$ for all $t\ge t_s$. The solution
of $\dot{q}=\velprofileso_2(q)$ through $(q(t_s^+),\dot{q}(t_s^-))$
satisfies $\ddot{q}=\velenergyso_2'(q)=\nominalacc_2(q,\velprofileso_2(q))$
by \emph{ii} of Def.~\ref{def:feedbackmap}, so it solves
$\ddot{q}=\nominalacc_2(q,\dot{q})$ from the same state, and
uniqueness, $\nominalacc_2$ being locally Lipschitz, gives the claim.
\end{proof}

Note that violating only the acceleration condition in
\eqref{eq:splice-both} gives a disturbance jump of
$(b-1)\Delta\nominalacc$ as considered in
Corollary~\ref{cor:fo-transient}, and by Theorem~\ref{thm:so-fo}
exact rejection is recovered in finite time for sufficiently large
$L$.

\begin{remark}
If $(\velprofileso_2,\feedbackmapso_2)$ is instead as in
Proposition~\ref{prop:so-fo-terminal}, then i) holds as stated.
Moreover, if $e_1(t_s^-)=0$, then so do ii) with $T=t_s$ and iii), by
case b) of that proposition, but now with $(q,\dot{q})$ reaching the
target in finite time.
\end{remark}

By Proposition~\ref{prop:replanning}, replanning (replacing
the active profile by a new one through the current state) can be achieved through the following steps:
\begin{enumerate}[label=\arabic*), leftmargin=*]
 \item construct a new profile $\velprofileso_2$ that passes through the current state
 according to \eqref{eq:splice-both}, as well as satisfying
 Def.~\ref{def:velocity-profile} and any speed and
 acceleration limits of the task;
 \item retain the old feedback map $\feedbackmapso_1$ if it satisfies
 Def.~\ref{def:feedbackmap} for $\velprofileso_2$, or choose a new
 $\feedbackmapso_2$ that does;
 \item retain the rejection pair $(\velprofilefo,\feedbackmapfo)$ and
 $L$, and $\injectionshape$ unless $e_1(t_s^-)=0$, as in
 Proposition~\ref{prop:replanning};
 \item switch to $(\velprofileso_2,\feedbackmapso_2,\injectionshape_2)$ at $t_s$, carrying over $\velestimate$ and $\distestimate$.
\end{enumerate}
Steps 2)--4) are immediate; step 1) is a planning problem, which we
now show can be posed as a convex  problem:

\subsubsection*{Planning as a convex program}
Consider step 1) in terms of $\velenergyso=\tfrac{1}{2}\velprofileso^2$.
Let $I$ be the closed interval between the switch position
$q_s:=q(t_s^+)$ and the target $q=0$, e.g., $I:=[q_s,0]$ if $q_s<0$.
Here we naturally assume $q_s\dot{q}(t_s^-)\le0$, i.e., that the
motion is toward the new target, as \eqref{eq:splice} is otherwise
not possible. Further, let $\dot{q}_s:=\dot{q}(t_s^-)$ and
$a_s:=\nominalacc_1(q(t_s^-),\dot{q}_s)$ be the velocity and nominal
acceleration at the switch, and $v_{\min},v_{\max}:I\to\R_{\ge0}$
continuous speed limits with $0<v_{\min}(q)\le v_{\max}(q)$ for
$q\neq0$ and $v_{\min}(q)\ge c|q|$ near the target for some $c>0$. Thus, following
\cite{verscheure2009time,lipp2014minimum}, the planning problem can
then be posed as a convex program:
\begin{subequations}
 \label{eq:replan}
 \begin{align}
  \min_{\velenergyso\in\mathcal{C}^{1,1}(I)}\quad
  &\int_I\ell\bigl(\velenergyso,\velenergyso'\bigr)\,dq
  \label{eq:replan-obj}\\
  \text{s.t.}\quad
  &\velenergyso(q_s)=\tfrac{1}{2}\dot{q}_s^2,\quad
  \velenergyso'(q_s)=a_s,
  \label{eq:replan-switch}\\
  &\velenergyso(0)=0,\quad \velenergyso'(0)=0,
  \label{eq:replan-adm}\\
  &\tfrac{1}{2}v_{\min}(q)^2\le\velenergyso(q)\le\tfrac{1}{2}v_{\max}(q)^2,
  \label{eq:replan-vel}\\
  &|\velenergyso'|\le a_{\max},\quad
  |\velenergyso''|\le\kappa_{\max},
  \label{eq:replan-lim}
 \end{align}
\end{subequations}
The constraints\footnote{Further convex constraints can be added: a fixed
final approach $\velenergyso=\velenergyso_1$ near the target; or, for
$\feedbackmapso(q,s)=\beta s$, condition \emph{iv} of
Def.~\ref{def:feedbackmap}, $|\velenergyso'|\le\beta'\sqrt{2\velenergyso}$
with $\beta'<\beta$, which settles step 2) within the plan.}
\eqref{eq:replan-switch}--\eqref{eq:replan-lim} are linear in
$\velenergyso$, so \eqref{eq:replan} is indeed a convex program for
any convex $\ell$. Here $|\velenergyso''|\le\kappa_{\max}$ bounds the jerk
$\velenergyso''\velprofileso$ along the profile by
$\kappa_{\max}v_{\max}$, the jerk limit itself being nonconvex. Since $\velenergyso'$
is the nominal input, a natural choice in \eqref{eq:replan-obj} is the
effort $\ell=\velenergyso'^2$, or an LQR-like cost à la $\ell=(\tfrac{1}{2}v_{\max}(q)^2-\velenergyso)^2+r\velenergyso'^2$
with $r>0$.\footnote{Under Theorem~\ref{thm:so-fo} the approach to the target is
asymptotic, so the time of arrival is finite only to a neighborhood
$\{|q|\ge\epsilon\}$ of it, $\epsilon>0$. This time,
$\int_{I\setminus\{|q|<\epsilon\}}dq/\sqrt{2\velenergyso}$, is convex
in $\velenergyso$, so it can be bounded as a further constraint or
minimized as the objective, with $\ell=1/\sqrt{2\velenergyso}$ (see  \cite{verscheure2009time,lipp2014minimum}).}

Given a solution $\velenergyso$, the new profile is simply
$\velprofileso_2(q)=-\sgn(q)\sqrt{2\velenergyso(q)}$ on $I$, admissible
there by \eqref{eq:replan-vel}, and extended beyond $I$ by an
admissible profile, matched in $\velenergyso$ and $\velenergyso'$ at
$q_s$. 

Of course, the limits in \eqref{eq:replan}
concern the nominal motion, and during a rejection transient,
$\ddot{q}=\nominalacc+\dot{\velestimateerror}$ differs from it by
$\dot{\velestimateerror}$, which Corollary~\ref{cor:pace-envelope}
bounds only the integral $\int|\dot{\velestimateerror}|\,dt$, not the
magnitude. Moreover, since all feasible plans share the
bounds \eqref{eq:replan-vel}--\eqref{eq:replan-lim} and the extension, and with them
the Lipschitz constant of $\nominalacc$, one $L\ge L^\star$ serves
them all if the feedback map is retained. If \eqref{eq:replan} is
infeasible, as when a new target is too close to decelerate within
$a_{\max}$, the current profile can simply be retained. It brings the motion to
rest at the old target, from where a new plan can be made. 

This type
of switching mechanism has two further use cases:

\textbf{1. Soft start:} A motion from rest at $q_s\neq0$ cannot start
on a profile, as Def.~\ref{def:velocity-profile} requires
$\velprofileso(q_s)\neq0$, while starting off the profile makes the
departure a transient of $\feedbackmapso$ with a jump in the
commanded acceleration. Since \eqref{eq:so-fo-controller-L} rejects
$\tilde{d}=(b-1)k+d$ for any bounded-rate command $k$, start-up can
instead use a time-based command (see also \cite{saetre2023orbital} for an alternative),
such as a jerk-limited acceleration ramp $k=a(t)$, until a speed
$v_s>0$ is reached, and then follow steps 1)--4) with $a_s=a(t_s)$.
As the rejection loop has converged during the ramp, the switch is
transient-free.

\textbf{2. Gain scheduling:} Since $\injectionshape$ and
$(\velprofilefo,\feedbackmapfo)$ enter only through the rejection
loop, they can be replaced at any instant with zero error, without
affecting the input or the motion, and so can $\feedbackmapso$ once
the profile is reached, as every admissible feedback map vanishes
there. Theorem~\ref{thm:so-fo} or
Proposition~\ref{prop:so-fo-terminal} then certifies the new set on
the same compact set. This allows, e.g.,
stiffer damping near the target or a higher $L$ where load changes
are expected, without the slow-variation conditions of conventional gain
scheduling (see, e.g., \cite{rugh2000research}).

\section{Example: Elevator control}\label{sec:example}
We now illustrate PACER and the planner of Section~\ref{sec:replanning}
in numerical simulation of a fitting application: an elevator. Indeed, an elevator must stop precisely at the intended floor within a given time, respect comfort limits on its speed, and handle loads that vary significantly from trip to trip. 

\begin{figure}[t]
 \centering
 \includegraphics{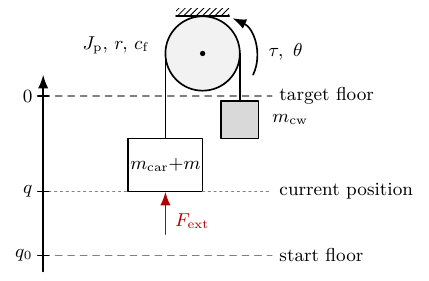}
 \caption{Schematic of the elevator system in Section~\ref{sec:example}.
 }
 \label{fig:elevator-schematic}
\end{figure}

Neglecting cable
elasticity, its dynamics can be modelled as a one-degree-of-freedom mechanical system:
\begin{equation}
 J(m)\ddot{\theta}=\tau-c_{\rm f}\dot{\theta}-\tau_{\rm L}+\tau_{\rm ext}(t),
 \label{eq:mech}
\end{equation}
where (see Fig.~\ref{fig:elevator-schematic})  $\theta$ is the angle of the traction pulley,  $\tau$ is the motor torque,
 $c_{\rm f}$ is the viscous friction coefficient, while $\tau_{\rm ext}=rF_{\rm ext}$ is a torque arising from the external force $F_{\rm ext}$ acting on the car  and the pulley radius $r$. Moreover,
\begin{align*}
J(m)&:=J_{\rm p}+(m_{\rm car}+m+m_{\rm cw}+m_{\rm cab})r^2,\\
 \tau_{\rm L}&=(m_{\rm car}+m-m_{\rm cw})g_0r,
\end{align*}
where $J_{\rm p}$ is the pulley's moment of inertia;
$m_{\rm car}$, $m_{\rm cw}$ and $m_{\rm cab}$ denote the mass of the car, counterweight and
cable, respectively;  $g_0$ is the gravitational acceleration; and
$m\in[0,\bar{m}]$ the unknown passenger mass, with
$\bar{m}=\SI{1000}{\kilogram}$.

Let $\theta^\star$ denote the pulley angle at the target floor, so that $q:=r(\theta-\theta^\star)$ is the car height relative to the target
floor. Then for $\tau=J(\bar{m})u/r$, \eqref{eq:mech} takes the form of
\eqref{eq:second-order} with
\begin{align*}
 b&=J(\bar{m})/J(m)\in[1,J(\bar{m})/J(0)],\\
 d&=\bigl(r(\tau_{\rm ext}(t)-\tau_{\rm L})-c_{\rm f}\dot{q}\bigr)/J(m).
\end{align*}

We will now apply PACER to a single upward trip from rest at $q(0)=q_0<0$.
Its three elements are chosen in turn: 1) the profile $\velprofileso$,
which sets the nominal motion (Def.~\ref{def:velocity-profile}); 2) the
feedback map $\feedbackmapso$, which corrects deviations from it
(Def.~\ref{def:feedbackmap}); and 3) the rejection pair
$(\velprofilefo,\feedbackmapfo)$ with scaling $\injectionshape$ and
gain $L$, which cancel the lumped disturbance
(Theorem~\ref{thm:so-fo}):

\begin{figure}[!t]
 \centering
 \includegraphics{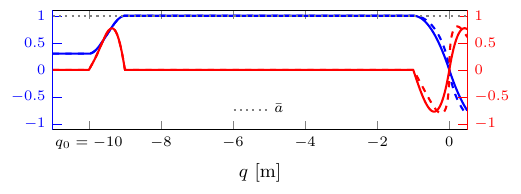}
 \caption{The profile $f$ from \eqref{eq:elevator-profile} (blue) and the nominal
 acceleration $\velenergyso'$ (red) used in the elevator example for the
 values in Table~\ref{tab:elevator-parameters}, for $\alpha=1$
 (solid) and $\alpha=0.75$ (dashed).}
 \label{fig:elevator-profile}
\end{figure}

\textbf{1) Profile:} The profile must cruise at the speed limit
$v_{\max}$, slow down smoothly over the last $h_{\rm d}$ before the
floor and, since Def.~\ref{def:velocity-profile} rules out
$\velprofileso(q_0)=0$, start at a creep speed and ramp up over the
first $h_{\rm a}$. We therefore take the product of a deceleration
curve $P$ and a ramp-up factor $R$:
\begin{equation}
 \velprofileso(q)=-v_{\max}\,P\Bigl(\frac{q}{h_{\rm d}}\Bigr)
 \Bigl(r_{\rm c}+(1-r_{\rm c})\,R\Bigl(\frac{q-q_0}{h_{\rm a}}\Bigr)\Bigr).
 \label{eq:elevator-profile}
\end{equation}
Here $r_{\rm c}\in(0,1)$ is the creep ratio, and $-q_0\ge h_{\rm a}+h_{\rm d}$;
$P(x):=\sgn(x)(2|x|^{\alpha}-|x|^{2\alpha})$ for $|x|\le1$, $P(x):=\sgn(x)$ otherwise, with $\alpha\in[\tfrac{2}{3},1]$;\footnote{For $\alpha<1$, the profile is not of the specific form required by \eqref{eq:general-homogeneity} and \eqref{eq:hom_pair}. However, we use it here as both can (given more space) be extended to this class using the method in \cite{bhat2005geometric}.}
while
  $R(x):=3x^2-2x^3$ for $x\in[0,1]$, $R(x):=0$ for $x<0$,
$R(x):=1$ for $x>1$. We use in the following both $\alpha=1$ (a parabola) and
$\alpha=0.75$ (see
Fig.~\ref{fig:elevator-profile}).

The deceleration $\velenergyso'$ is bounded, and so is
the jerk $\velenergyso''\velprofileso$ along the profile, which
behaves as $|q|^{3\alpha-2}$ at the floor.

\textbf{2) Feedback map:} Inspired by the optimal
nonlinear damping of \cite{ruderman2021optimal}, we take the feedback
map of Def.~\ref{def:feedbackmap} as
\begin{equation}
 \feedbackmapso(q,s)=\bar{a}\tanh\Bigl(\frac{1}{\bar{a}}\Bigl(\beta s+\gamma\,\frac{s|s|}{|q|+\Delta_q}\Bigr)\Bigr),
 \label{eq:elevator-Gamma}
\end{equation}
for the  $\alpha=1$ profile, 
with $\beta>2v_{\max}/h_{\rm d}$, $\gamma\ge0$, $\Delta_q>0$ and
$\bar{a}>0$. Deviations are thus corrected increasingly firmly as the
floor approaches, with the corrective acceleration bounded by
$\bar{a}$, and Def.~\ref{def:feedbackmap} holding for $\bar{a}$
sufficiently large. 

For the (terminal) $\alpha=0.75$  profile, the pair \eqref{eq:hom_pair} is
required exactly near the floor, so we take instead
\begin{equation*}
 \feedbackmapso(q,s)=
 \begin{cases}
 \beta\spow{s}{\frac{2\alpha-1}{\alpha}},  \quad \text{when} \quad  \beta|s|^{\frac{2\alpha-1}{\alpha}}\le\tfrac{1}{2}\bar{a},
 \\[2pt]
 \tfrac{1}{2}\bar{a}\,\sgn(s)\Bigl(1+\tanh\Bigl(\tfrac{2\beta}{\bar{a}}|s|^{\frac{2\alpha-1}{\alpha}}-1\Bigr)\Bigr) \ \text{otherwise},
 \end{cases}
\end{equation*}
with the same $\beta$ and $\bar{a}$. This is continuous, $\mathcal{C}^1$ away from $s=0$, equal
to the map in \eqref{eq:hom_pair} below $\bar{a}/2$
and saturating at $\bar{a}$, and $\beta>\alpha(2v_{\max}/h_{\rm d}^{\alpha})^{1/\alpha}$
holds. We refer in the following to this finite-time profile design as PACER-FT, and to the one with
$\alpha=1$ and \eqref{eq:elevator-Gamma} simply as PACER.

\textbf{3) Disturbance rejection:} We use the pair
\begin{subequations}
    \label{eq:elevator-rejection}
\begin{align}
 \velprofilefo(\velestimateerror)&=-\kappa_{\rm e}\spow{\velestimateerror}{1/2},
 \label{eq:elevator-g}\\
 \feedbackmapfo(\velestimateerror)&=
 \begin{cases}
 \mu_{\rm e}\spow{\velestimateerror}{1/2},
 & |\velestimateerror|\le\bar{y},\\
 \mu_{\rm e}\sgn(\velestimateerror)
 \Bigl(\sqrt{\bar{y}}+\dfrac{|\velestimateerror|-\bar{y}}{2\sqrt{\bar{y}}}\Bigr),
 & |\velestimateerror|>\bar{y},
 \end{cases}
 \label{eq:elevator-nu}
 \end{align}
 \end{subequations}
with $\mu_{\rm e}>\kappa_{\rm e}>0$ and $\bar{y}>0$. Near zero this is the standard super-twisting pair
\cite{levant1993sliding}, so \eqref{eq:fo-local-terminal} holds with
$\alpha=\tfrac12$ and $\injectionbound=\kappa_{\rm e}(\mu_{\rm e}-\kappa_{\rm e})/2$
in \eqref{eq:fo-rate-bound}. Beyond $\bar{y}$, $\feedbackmapfo$ is linear,
so large errors are damped proportionally rather than by the square
root, while $\velprofilefo$ is unchanged. The pair is thus not proportional and lies outside the
generalized super-twisting family of \cite{moreno2011lyapunov}. 

We also
use the scaling function
\begin{equation}
 \injectionshape(q)=1+(\bar{\omega}-1)\bigl(1-e^{-(q/\Delta_\omega)^2}\bigr),
 \label{eq:elevator-omega}
\end{equation}
$\bar{\omega}\ge1$, $\Delta_\omega>0$, which reduces the gain near the
floor, where measurement noise matters most.

\begin{table}[!t]
 \centering
 \footnotesize
 \caption{Parameters used in the elevator example.}
 \label{tab:elevator-parameters}
 \begin{tabular}{@{}l@{}}
 \toprule
 Plant: $(m_{\rm car},m_{\rm cw},m_{\rm cab},r,J_{\rm p},c_{\rm f})=(1000,1500,200,0.4,10,16)$\\
 Unmodelled: $(F_{\rm c},v_{\rm c},T_{\rm a},n_{\rm e})=(200,0.01,0.005,10^{-6})$\\
 External force: $(F_{\rm p},F_{\rm b},T_{\rm b},t_{\rm b})=(500,2000,1,5)$\\
 Profile \eqref{eq:elevator-profile}: $(v_{\max},h_{\rm a},h_{\rm d},r_{\rm c})=(1,1,1,0.3)$\\
 Feedback map \eqref{eq:elevator-Gamma}: $(\beta,\gamma,\Delta_q,\bar{a})=(3,1,0.5,1)$\\
 Rejection \eqref{eq:elevator-rejection}--\eqref{eq:elevator-omega}: $(\kappa_{\rm e},\mu_{\rm e},\bar{y},\bar{\omega},\Delta_\omega,L)=(1,3,0.05,1.83,1.5,4)$\\
 \bottomrule
 \end{tabular}
\end{table}

\textbf{Simulation setup:} The numerical simulations were carried out in
MATLAB. The controllers were run at \SI{1}{\kilo\hertz}, with their dynamic
parts discretized by forward Euler and the velocity obtained as the
backward difference of the sampled position; between samples, the \texttt{ode45} solver was used with relative and absolute
tolerances of $10^{-8}$ and $10^{-10}$, respectively, with the input held
constant (zero-order hold). 

In the replanning
problem \eqref{eq:replan}, $\velenergyso$ was taken as a $\mathcal{C}^1$
quadratic spline with knots $q_i$ spaced \SI{2}{\centi\metre} apart,
i.e., $\velenergyso''(q)=w_i$ on $[q_i,q_{i+1})$, and the cost
integrated exactly using a three-point Gauss quadrature. The resulting
quadratic program in the $w_i$ was solved with the \texttt{quadprog}
solver from the Optimization Toolbox.

PACER was initialized with
$\velestimate(0)=\dot{q}(0)$ and $\distestimate(0)=0$. To test the
design beyond its assumptions, the model also included Coulomb
friction, smoothed as $F_{\rm c}\tanh(\dot{q}/v_{\rm c})$, whose
Lipschitz constant $F_{\rm c}/v_{\rm c}$ far exceeds what $L$ was
sized for; an actuator lag $T_{\rm a}$, which is unmodelled
dynamics; and a small Gaussian encoder noise of standard deviation
$n_{\rm e}=\SI{1}{\micro\metre}$. The parameters are given in
Table~\ref{tab:elevator-parameters} (all in SI units); with these,
$\beta>2v_{\max}/h_{\rm d}$ and $\bar{a}>\max_q|\velenergyso'(q)|$,
so Def.~\ref{def:feedbackmap} holds, and $L=4$ is suggested by
\eqref{eq:L-rule-of-thumb} with $\delta\approx 1$, $\Gamma=1$,
$\sup|\dot{k}|=|\velenergyso''\velprofileso|_{\max}=4.2$ and
$\sup|\dot{d}|=2.4$ from the rate of $F_{\rm ext}$, which gives
$L\approx3.9$.

We consider here three simulated scenarios:
\begin{enumerate}[leftmargin=*]
\item A trip from $q_0=\SI{-10}{\metre}$ with a passenger mass of
 $m=\SI{250}{\kilogram}$, for which $b=1.25$, and the external force
 $F_{\rm ext}(t)=F_{\rm p}\sin(\pi t/2)+F_{\rm t}(t)$, where
 $F_{\rm t}(t)=-F_{\rm b}\sin^2\!\bigl(\pi(t-t_{\rm b})/T_{\rm b}\bigr)$
 for $t\in[t_{\rm b},\,t_{\rm b}+T_{\rm b})$ and $F_{\rm t}(t)=0$
 otherwise (its rate satisfies the bound of Theorem~\ref{thm:so-fo}).

\item A restart from rest at $q=\SI{-8}{\metre}$, i.e., \SI{1}{\metre\per\second} off the
 profile, without the soft start of Section~\ref{sec:replanning},
 with the same car and the force $F_{\rm p}\sin(\pi t/2)$.

\item A new call during the trip of Scenario 1 without $F_{\rm t}$:
 when the car reaches $q=\SI{-0.6}{\metre}$ during deceleration, the target
 floor is moved from $0$ to \SI{5}{\metre}. The new profile is the solution
 of \eqref{eq:replan} with the LQR-like cost
 $\ell(\velenergyso,\velenergyso')=\velenergyso'^2+(\velenergyso-\tfrac{1}{2}v_{\max}^2)^2$,
 $v_{\min}=r_{\rm c}v_{\max}$ on $I$, $a_{\max}=0.8$,
 $\kappa_{\max}=j_{\max}/v_{\max}$ with $j_{\max}=1$,  and, outside these limits, the deceleration segment of
 \eqref{eq:elevator-profile} retained as the final approach, switched in with and without the switch conditions
 \eqref{eq:splice-both}.

\end{enumerate}

For comparison, we also simulated the following controllers, where $e:=q_r-q$ with $q_r(t)$ denoting a seven-segment S-curve
reference with jerk and acceleration limits of \SI{1}{m/s^3} and
\SI{0.8}{m/s^2}, and the same speed limit as PACER:
\begin{itemize}[leftmargin=*]
\item Prop.~\ref{prop:so-dyn}'s integral action (IA):
 $u=\nominalacc-\lambda\velestimateerror$, i.e., PACER without
 $\distestimate$, with $\lambda=20$ (the largest gain at which its
 chattering stays at the level of the linear designs below).
 
\item S-curve + PID: $u=\ddot{q}_{\rm r}+3\omega_{\rm p}^2e+3\omega_{\rm p}\dot{e}+\omega_{\rm p}^3\!\int_0^t e\,d\tau$,
 $\omega_{\rm p}=5$.

\item S-curve + ADRC \cite{han2009pid,gao2003scaling}:
 $u=\ddot{q}_{\rm r}+\omega_{\rm c}^2(q_{\rm r}-z_1)+2\omega_{\rm c}(\dot{q}_{\rm r}-z_2)-z_3$,
 where $\dot{z}_1=z_2-3\omega_{\rm o}\tilde{e}$,
 $\dot{z}_2=z_3+u-3\omega_{\rm o}^2\tilde{e}$,
 $\dot{z}_3=-\omega_{\rm o}^3\tilde{e}$, $\tilde{e}:=z_1-q$,
 $\omega_{\rm c}=5$, $\omega_{\rm o}=30$, i.e., with nominal input gain
 $b_0=1$ as for PACER.

\item VFC\,+\,PI:
 $u=\velprofileso'(q)\dot{q}+2\omega_{\rm v}\velerrorso+\omega_{\rm v}^2\!\int_0^t\velerrorso\,d\tau$,
 $\omega_{\rm v}=5$, i.e., the cancellation $\velprofileso'(q)\dot{q}$
 gives linear PI dynamics in $\velerrorso$ (cf.
 velocity field control (VFC) \cite{li1999passive,cervantes2002robust}).
\item VFC\,+\,STA:
 $u=\velprofileso'(q)\dot{q}+\sqrt{L_{\rm sta}}\,\mu_{\rm e}\spow{\velerrorso}{1/2}+w$,
 $\dot{w}=\tfrac12L_{\rm sta}\kappa_{\rm e}(\mu_{\rm e}-\kappa_{\rm e})\sgn(\velerrorso)$,
 with $L_{\rm sta}=2$ (chosen so that its speed deviation while
 $F_{\rm t}$ acts is comparable to that of PACER), i.e., using the super-twisting algorithm
 \cite{shtessel2014sliding}.
\end{itemize}
The linear controllers thus share a closed-loop bandwidth of
\SI{5}{\radian\per\second} and were not retuned for $F_{\rm t}$. All were used in
Scenario 1, with VFC\,+\,PI and VFC\,+\,STA also used in Scenario 2.

\begin{figure}[!t]
 \centering
 \includegraphics{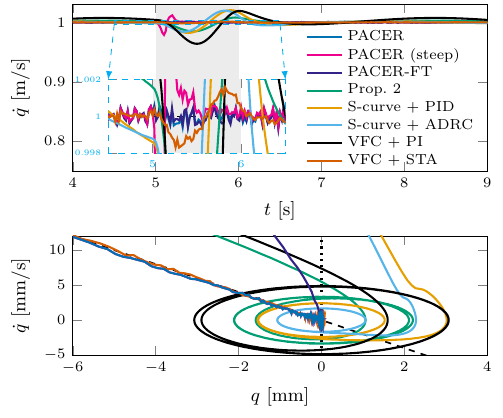}
 \caption{Speed and arrival of the car in Scenario 1 under PACER
 \eqref{eq:so-fo-controller-L}, PACER-FT and the comparison controllers. Top:
 speed while the transient load $F_{\rm t}$ acts (shaded interval).
 Bottom: phase plane of the arrival at the floor, with the $\alpha=1$ profile
 dashed; the designs with linear integral action settle on limit
 cycles, while PACER and VFC\,+\,STA settle within \SI{0.1}{\milli\metre} of the
  intended floor.}
 \label{fig:elevator-trip}
\end{figure}

\textbf{Results:} As shown in Fig.~\ref{fig:elevator-trip} for
Scenario 1, PACER keeps the speed within \SI{0.1}{\percent} of $v_{\max}$ while
$F_{\rm t}$ acts and settles within \SI{0.1}{\milli\metre} of the floor,
PACER-FT likewise but about \SI{3}{\second} earlier and without
overshoot; VFC\,+\,STA does the same within \SI{0.2}{\percent}. The other designs deviate by
\SIrange{1.5}{3.6}{\percent} in speed, overshoot by \SIrange{2}{3}{\milli\metre}
and converge to a limit cycle due to the dry friction. However, after
convergence, PACER (and VFC\,+\,STA) exhibit chatter. For PACER, this falls with the
sampling period, i.e., it is caused by the sampled discontinuous integrator, each sample being a break in the
sense of Corollary~\ref{cor:fo-transient}.

To test the rate bound of Theorem~\ref{thm:so-fo}, PACER was also
run with a steeper pulse of bounded variation, $F_{\rm b}=\SI{e4}{\newton}$ and
$T_{\rm b}=\SI{0.2}{\second}$, which violates it (magenta line); one can see that the speed deviates by \SI{2}{\percent} before the
profile is recovered within \SI{0.1}{\second}, which is consistent with Corollary~\ref{cor:fo-transient}.

\begin{figure}[!t]
 \centering
 \includegraphics{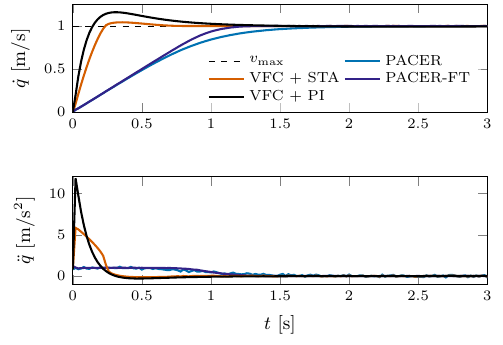}
 \caption{Restart  from rest (Scenario 2). Top: speed.
 Bottom: acceleration.}
 \label{fig:elevator-restart}
\end{figure}

As shown in Fig.~\ref{fig:elevator-restart} for Scenario 2, PACER and
PACER-FT reach the profile in \SIlist{1.5;1.3}{\second}, respectively,
both with a peak acceleration of \SI{1.6}{\metre\per\square\second}, set
by $\bar{a}$ of the feedback map and the gain $b$, whereas VFC\,+\,STA and
VFC\,+\,PI reach it in \SI{0.3}{\second} with peaks of
\SIlist{5.8;12}{\metre\per\square\second} and overshoot the speed limit,
since their single gain must both enforce the profile and reject the
disturbance. The restart of PACER could be made smoother still with
the soft start proposed in Section~\ref{sec:replanning}.

\begin{figure}[!t]
 \centering
 \includegraphics{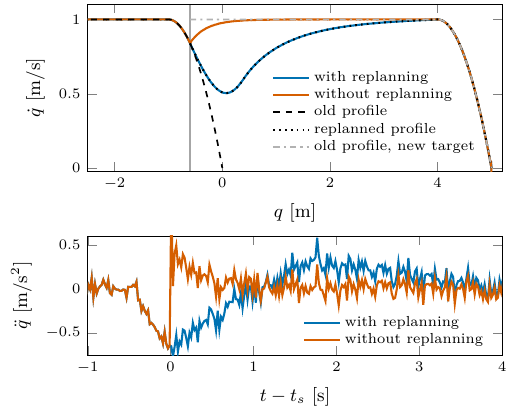}
 \caption{Response of the elevator under PACER to a new call during
 deceleration (Scenario 3), with and without the replanning scheme of
 Section~\ref{sec:replanning}: at $q=\SI{-0.6}{\metre}$ (grey line) the target
 floor is moved from $0$ to \SI{5}{\metre}. Top: phase plane with the old,
 replanned and shifted profiles, with the latter using the old profile for the new target. Bottom: acceleration after the call (with encoder noise clearly visible).}
 \label{fig:elevator-replan}
\end{figure}

As shown in Fig.~\ref{fig:elevator-replan} for Scenario 3, the car
follows the replanned profile from the switch on, and its
acceleration is continuous. Without replanning, the car must first catch up with the shifted profile, where the nominal acceleration $\nominalacc$ jumps from \SI{-0.67}{\metre\per\square\second} to \SI{0.45}{\metre\per\square\second} within one sample, and the car arrives \SI{1.1}{\second} earlier. Similar results (omitted for the
sake of space) were obtained with PACER-FT.

\section{Discussion and related work}
The approach in this paper rests on two elements: a velocity profile
$\dot{q}=\velprofileso(q)$, and the nominal input $\velenergyso'$ on
it, with $\velenergyso:=\tfrac12\velprofileso^2$. Deviations
$\velerrorso=\velprofileso(q)-\dot{q}$ are corrected by a feedback
map $\feedbackmapso$ of Def.~\ref{def:feedbackmap}, which gives
convergence for every such pair (Thm.~\ref{thm:general-f}). Since
the controller uses $\velenergyso'$ rather than cancelling the profile
dynamics $\velprofileso'\dot{q}$ along the actual motion, any
admissible $\feedbackmapso$ may be used, and terminal profiles need
no reparameterization (cf.\ \cite{feng2002non}). Importantly, by
anchoring the control to the profile in this way, the gain mismatch
and the perturbation lump into one disturbance, which the loop of
Section~\ref{sec:so-dyn} rejects exactly (Thm.~\ref{thm:so-fo}).
This is what we called profile-anchored control with exact rejection
(PACER), the main contribution in this paper: the nominal motion is
recovered under an unknown gain and perturbation. Since the profile is
anchored to position, we can moreover replace it along the way by a
new one through the current state, planned in $\velenergyso$ itself
(Sec.~\ref{sec:replanning}).

The rejection loop of PACER is familiar from disturbance observers
\cite{ohishi1987microprocessor,chen2016disturbance}, including in
super-twisting form \cite{davila2005second}, and integral sliding
modes \cite{utkin1996integral,castanos2006analysis}, and, as in
active disturbance rejection and model-free control
\cite{han2009pid,fliess2013model}, the unknown gain is absorbed into
the disturbance. 
What
Theorem~\ref{thm:so-fo} adds is that the nominal motion is recovered
exactly after a finite time, semiglobally and for any
$b\in[\underline{b},\bar{b}]$, where the performance recovery of
\cite{freidovich2008performance,back2008adding} is approximate, and
where the sliding-mode observer of \cite{kim2025sliding} recovers a
nominal loop exactly in finite time as well, but for linear plants
with a fixed baseline controller. The
envelope of Corollary~\ref{cor:envelope} is then kept up to a margin
of order $L^{-1/2}$ (Corollary~\ref{cor:pace-envelope}), and, for the
finite-time designs, exactly after a finite time
(Prop.~\ref{prop:so-fo-terminal}).

Velocity field control
\cite{li1999passive,cervantes2002robust} is perhaps the nearest neighbour on
the motion-control side; with the same pair it rejects the load
equally well (Scenario 1). However, it places no bound on the
correction (Scenario 2), and since its integrator acts on $\velerrorso$ rather than on
$\velestimateerror$, it allows the switch of
Section~\ref{sec:replanning} only once the profile is reached, not
during a transient (cf.\ Prop.~\ref{prop:replanning}). Such switching without a transient is the other trait of PACER. With
the disturbance rejected and its estimate carried over, a new target,
a changed speed limit or an obstacle is handled by solving a convex
program once (Sec.~\ref{sec:replanning}), and in between the plan
can be left alone. Handling such events online is otherwise the
domain of optimization-based control, where related schemes, such as
path-following model predictive control
\cite{faulwasser2016nonlinear}, reference governors
\cite{garone2017reference} and online trajectory generation
\cite{kroger2010online}, re-solve at every sample in their standard
form.

On its own, the rejection loop of PACER is the dynamic controller
\eqref{eq:fo-dyn-controller} for first-order systems. Depending on
the pair $(g,\nu)$, it ranges from PI control to the super-twisting
algorithm and its disturbance-tailored variants
\cite{haimovich2019disturbance}. Compared with the Lyapunov framework
of \cite{moreno2011lyapunov}, it admits non-proportional pairs, for
which $\dot{\xi}=\tfrac{1}{2}(\velprofilefo^2)'+\velprofilefo\feedbackmapfo'$
is not of the form $-k_2\phi\phi'$, and its disturbance class is set
directly by the rate bound $\delta\Gamma$, which scaling the pair
meets. The price is a restricted gain set, where the standard STA pair has gains
$\mu$ and $\kappa(\mu-\kappa)/2\le\mu^2/8$, excluding the common
setting $(1.5\sqrt{C},1.1C)$ for $|\dot{d}|\le C$
\cite{levant1998robust,seeber2017stability}, while the PI pair has a
damping ratio of at least $\sqrt{b}$.

Some limitations of the presented scheme nevertheless remain:
Theorem~\ref{thm:so-fo} requires a locally Lipschitz nominal
feedback, and Proposition~\ref{prop:so-fo-terminal} lifts this only
for the pair \eqref{eq:hom_pair}, imposed near the target and along
the profile; general homogeneous pairs would follow the same route.

On the implementation side, we have to a large extent neglected
things like input saturation, actuator dynamics, delays, and
measurement noise. Saturation, and consequently windup, is especially
delicate due to the integrator $\distestimate$ in
\eqref{eq:so-fo-controller-L}, although with the super-twisting pair
its rate is bounded, so
anti-windup schemes such as back-calculation
\cite{astrom1989integrator} should be viable. Delays and actuator dynamics, on the other hand, bound the usable
$L$ from above, as in any high-gain loop, while measurement noise can be mitigated by lowering the injection
gain through $\injectionshape$ where it matters, as done near the floor in the example.

Lastly, the design is carried out in continuous-time. Thus, a sampled
implementation generally alters the finite-time properties, and the chatter in
the example is caused by the sampled discontinuous integrator, each
sample being a break in the sense of
Corollary~\ref{cor:fo-transient}. This motivates the use of an
implicit discretization method à la \cite{brogliato2021digital},
which might reduce the variation per sample and thus chatter.

\section{Conclusion}
We have shown that a position-dependent velocity profile can serve
both as the specification of a motion and as the nominal input of the
controller that executes it. The resulting profile-anchored controller with exact rejection, 
coined PACER, recovers the nominal motion exactly after a finite time (even under
time-varying perturbations and an unknown input gain); moreover, it
allows (re-)planning a new profile under speed, acceleration and
jerk limits via a convex program, and the switch can be carried out
without a transient. Extending this to several degrees of freedom along a
path, and testing it on hardware, are the next steps.

\begingroup
\small
\bibliographystyle{IEEEtran}
\bibliography{references.bib}
\endgroup
\end{document}